\newcommand{\alt}{\mathrel{\raise.3ex\hbox{$<$\kern
-.75em\lower1ex\hbox{$\sim$}}}}
\newcommand{\agt}{\mathrel{\raise.3ex\hbox{$>$\kern
-.75em\lower1ex\hbox{$\sim$}}}}

\documentclass[varenna]{cimento}

\usepackage{epsfig}  
\usepackage{color}  
\usepackage{amsmath}
\usepackage{mathbbol}

\title{Exact numerical methods for electron-phonon problems} 

\author{Eric Jeckelmann}

\institute{Institut f\"{u}r Theoretische Physik\\
Universit\"{a}t Hannover\\
Appelstra\ss e 2, D-30167 Hannover, Germany}

\author{Holger Fehske}

\institute{Institut f\"{u}r Physik\\
Ernst-Moritz-Arndt-Universit\"{a}t Greifswald\\
D-17487 Greifswald, Germany}

\begin{document}

\maketitle

\section{Introduction \label{sec:intro}}

In the last few years solid state physics has increasingly benefited from
scientific computing and the significance of numerical techniques 
is likely to keep on growing quickly in this field. 
Because of the high complexity of solids, which  are made of a huge number 
of interacting electrons and nuclei, a full understanding 
of their properties cannot be developed using analytical methods only.
Numerical simulations do not only provide quantitative results for 
the properties of specific materials but are also widely used
to test the validity of theories and analytical approaches.

Numerical and analytical approaches based on perturbation theory
and effective independent-particle theories such as
the Fermi liquid theory, the density functional theory,
the Hartree-Fock approximation, or the Born-Oppenheimer approximation, 
have been extremely successful in explaining the properties of
solids.  
However, the low-energy and low-temperature electronic, optical, or 
magnetic properties
of various novel materials are not understood within these simplified
theories.
For example, in strongly correlated
systems, the interactions between constituents of the solid are
so strong that they can no longer be considered separately and
a collective behavior can emerge.
As a result, these systems may exhibit new and
fascinating macroscopic properties such as high-temperature superconductivity or
colossal magneto-resistance~\cite{science00}.
Quasi-one-dimensional  electron-phonon  (EP) systems like
MX-chain compounds are other examples of electronic systems that are
very different from traditional ones~\cite{MX93}.  
Their study is particularly rewarding for a number of reasons.  First they
exhibit a remarkably wide range of competing forces,
which gives rise to a rich variety of different phases
characterized by symmetry-broken ground states and
long-range orders.
Second these systems share fundamental features with higher-dimensional
novel materials (for instance, 
high-$T_{\rm c}$ cuprates or charge-ordered nickelates)  such as
the interplay of charge, spin, and lattice degrees of
freedom. 
One-dimensional (1D) models allow us to investigate 
this complex interplay, which is important but poorly understood 
also in 2D- and 3D  highly correlated electron systems, 
in a context more favorable to numerical simulations. 

\subsection{Models}

Calculating the low-energy low-temperature properties of solids
from first principles 
is possible only with various approximations which
often are not reliable in strongly correlated or low-dimensional
electronic systems. An alternative approach for investigating these materials
is the study of simplified lattice models which include only the relevant 
degrees of freedom and interactions but nevertheless are believed to  
reproduce the essential physical properties of the full system.

A fundamental model for 1D correlated electronic systems is the 
Hubbard model~\cite{Hubbard} defined by the Hamiltonian
\begin{equation}
H_{\rm ee} =- t \sum_{\langle i,j \rangle;\sigma} 
\left( c_{i\sigma}^{\dag}c^{\phantom{\dag}}_{j\sigma}
+ c_{j\sigma}^{\dag}c^{\phantom{\dag}}_{i\sigma} \right) 
+ U \sum_{i} n_{i\uparrow} n_{i\downarrow}  
.
\label{hubbard}
\end{equation}
It describes electrons with spin
$\sigma=\uparrow,\downarrow$
which can hop between neighboring sites on a lattice. 
Here $c^{\dag}_{i\sigma}$,
$c^{\phantom{\dag}}_{i\sigma}$ are creation and annihilation operators for
electrons with spin $\sigma$ at site $i$, $n_{i\sigma}=
c^{\dag}_{i\sigma}c^{\phantom{\dag}}_{i\sigma}$ 
are the corresponding density operators.
The hopping integral $t$ gives rise to a
a single-electron band of width $4tD$
(with $D$ being the spatial dimension). 
The Coulomb repulsion between electrons is
mimicked by a local Hubbard interaction $U \geq 0$.
The average electronic density per site is $0 < n < 2$,
where $n = N_{\rm e}/N$, $N$ is the number of lattice sites and 
$N_{\rm e}$ is the number of electrons; 
$n/2$ is called the band filling.

In the past the Hubbard model was intensively studied with respect to 
(itinerant) ferromagnetism, antiferromagnetism  
and metal-insulator (Mott) transitions in transition metals.
More recently it has been used in the context of heavy fermions and
high-temperature superconductivity as perhaps the most fundamental model
accounting for strong electronic correlation effects in solids. 

\begin{figure}[t]
\begin{center}
\includegraphics[width=13.0cm]{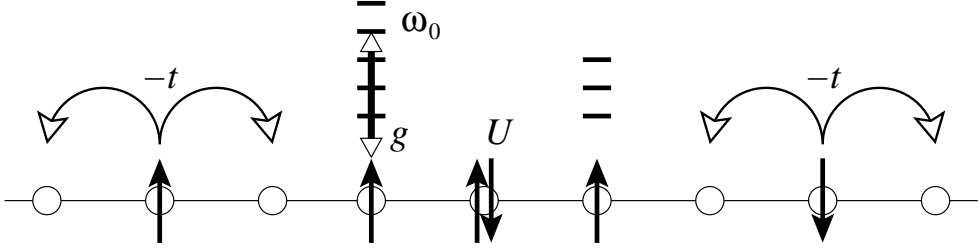}
\caption{Schematic representation of the 1D
Holstein-Hubbard model.   
}
\label{fig:model}
\end{center}
\end{figure}

The coupling between electrons and the lattice relaxation 
and vibrations (phonons) is also known to have significant 
effects on the properties of solids including the above mentioned
strongly correlated electronic materials
(see the papers by Egami, Calvani, Zhou, Perroni,
and Saini).
Dynamical phonon effects are particularly important
in quasi-1D metals and charge-density-wave (CDW) systems.  
The simplest model describing the effect of an additional EP
coupling is the Holstein-Hubbard model.
This model describes electrons coupled to dispersionless phonons,
represented by local Einstein oscillators. 
The  Hamiltonian is given by 
\begin{equation}
H_{\rm ep} = H_{\rm ee} + 
\frac{1}{2M}  \sum_i p^2_i  +  \frac{K}{2} \sum_i q^2_i
- \alpha  \sum_i  q_i  n_i  ,
\label{holstein}
\end{equation}
where $q_i$ and $p_i$ are the position and momentum operators
for a phonon mode at site $i$, and $n_i=n_{i\uparrow}+n_{i\downarrow}$.
At first sight, there are three additional parameters in this model
(compared to the purely electronic model): The oscillator
mass $M$, the spring constant $K$, and the EP coupling constant $\alpha$.
However, if we introduce phonon (boson) creation and annihilation operators 
$b^\dag_i$ and $b^{\phantom{\dag}}_i$, respectively,
the Holstein-Hubbard Hamiltonian can be written (up to a constant term)
\begin{equation}
H_{\rm ep} =  H_{\rm ee} + \omega_{0}  
\sum_i b^\dag_i b^{\phantom{\dag}}_i
- g \omega_0  \sum_i (b^\dag_i + b^{\phantom{\dag}}_i) 
n_i^{\phantom{\dag}},
\end{equation}
where the phonon frequency is given by 
$\omega_{0}^2 = K/M$ ($\hbar = 1$)
and a dimensionless EP coupling constant is defined by 
$g = \alpha a / \omega_0$
with the range of zero-point fluctuations
given by $2a^2 = (KM)^{-1/2}$. 
We can set the parameter $a$ equal to 1 by
redefining the units of oscillator displacements.
Thus, the effects of the EP coupling are determined
by two dimensionless parameter ratios only: $\omega_{0}/t$ and $g$.
Alternatively, one can use the polaron binding energy 
$\varepsilon_{\rm p} = g^2 \omega_{0}$ or, 
equivalently, $\lambda=\varepsilon_{\rm p}/2tD$, 
instead of $g$.
The various constituents and couplings of the Holstein-Hubbard model
are summarized in fig.~\ref{fig:model}.

In the single-electron case, the resulting Holstein model~\cite{Holstein} has
been studied as a paradigmatic model for polaron formation~(see the paper
by Fehske, Alvermann, Hohenadler and Wellein).  At half-filling, the EP
coupling may lead to a Peierls instability related to the 
appearance of CDW order in competition with the SDW 
instability triggered by $U$ (see the separate paper
by Fehske and Jeckelmann). 

\subsection{Methods}

Despite the great simplification brought by the above models,
theoretical investigations remain difficult
because a quantum many-particle problem has to be solved
with high accuracy to determine correlation effects 
on physical properties beyond the mean-field level.
Analytical solutions of these models are known for special
cases only.
To determine the spectral and thermodynamical properties
of these models, theorists have turned to numerical simulations.      
Among the various approaches, exact numerical methods
play a significant role.
A numerical calculation is said to be exact if
no approximation is involved aside from the restriction imposed 
by  finite computational resources
(in particular, the same calculation would be mathematically exact if
it were carried out analytically), 
the accuracy can be systematically improved
with increasing computational effort, and
actual numerical errors are quantifiable or
completely negligible. 
Especially for strongly correlated systems, 
exact numerical methods are often the only approach available to obtain 
accurate quantitative results in model systems and 
the results that they provide are essential for checking the
validity of theories or testing the accuracy of approximative 
analytical methods.  
Nowadays, finite-cluster exact diagonalizations (ED), the numerical
renormalization group (NRG), density
matrix renormalization group (DMRG) calculations, quantum Monte
Carlo (QMC) simulations, and the dynamical mean-field theory (DMFT)
have become very powerful and important tools for
solving many-body problems.  
In what follows, we briefly summarize the advantages and weaknesses of 
these techniques, especially in the context of  EP lattice models
such as the the Holstein-Hubbard model.
Then in the remaining sections we will present the basic principles
of the ED (sect.~\ref{sec:ed}) and DMRG (sect.~\ref{sec:dmrg}
and~\ref{sec:dynamical}) approaches.

The paradigm of an exact numerical calculation in a quantum system
is the exact diagonalization (ED) of its Hamiltonian,
which can be carried out using several
well-established algorithms (see the next section). 
ED techniques can be used to calculate most properties of any quantum system.
Unfortunately, they are restricted to small systems 
(for instance, 16 sites for a half-filled Hubbard model and 
less for the Holstein-Hubbard model) because of the exponential increase
of the computational effort with the number of particles.
In many cases, these system sizes are too small to simulate
adequately macroscopic properties of  solids (such as (band) transport, 
magnetism or charge long-range order). They are of great value,  
however, for a description of local or short-range order effects.

In QMC simulations the problem of computing quantum properties is transformed
into the summation of a huge number of classical variables, which is
carried out using statistical (Monte Carlo) techniques.
There are numerous variations of this principle also for EP problems
(see, e.g., the paper on QMC by Mishchenko).
QMC simulations are almost as widely applicable as ED techniques
but can be applied to much larger systems.  
The results of QMC calculations are affected by statistical errors
which, in principle, can be systematically reduced with increasing
computational effort. Therefore, QMC techniques are often 
numerically exact.
In practice, several problems such as the sign problem (typically,
in frustrated quantum systems) or the large auto-correlation time
in the Markov chain (for instance, in critical systems) severely
limit the applicability and accuracy of QMC simulations in
strongly correlated or low-dimensional systems.  
Moreover, real-frequency dynamical properties often
have to be calculated from the imaginary-time correlation functions
obtained with QMC using an analytic continuation.
However, the transformation of imaginary-time data affected
by statistical errors to real-frequency data
is an ill-conditioned numerical problem.
In practice, one has to rely on approximate transformations using
least-square or maximum-entropy fits,
which yield results of unknown accuracy.

DMRG methods allow us to calculate the static and dynamical
properties of quantum systems much larger than those
possible with ED techniques (see the third and fourth section, as well as 
the separate paper by Fehske and Jeckelmann).
For 1D systems and quantum impurity problems, 
one can simulate lattice sizes
large enough to determine static properties in the thermodynamic limit
and the dynamical spectra of macroscopic systems exactly.
In higher dimensions DMRG results are numerically exact
only for system sizes barely larger than those available with
ED techniques. For larger system sizes in dimension two and higher
DMRG usually provides only a variational approximation of the system 
properties.

The NRG is a precursor of the DMRG method. Therefore, it it
not surprising that most NRG calculations can also
be carried out with DMRG. NRG provides numerically exact results for 
the low-energy properties
of quantum impurity problems (see the paper on NRG methods by
Hewson). Moreover, for this type
of problem it is computationally more efficient than DMRG.
For other problems (lattice problems, high-energy properties) 
NRG usually fails or provides results of poor accuracy. 

In the dynamical mean-field theory 
(see the papers by Ciuchi, Capone, and Castellani)
it is assumed that the self-energy of the
quantum many-body system is momentum-independent. While this is exact on a
lattice with an infinitely large coordination number (i.e., in the
limit of infinite dimension), it is considered  to be a
reasonable approximation for 3D systems.
Thus in applications to real materials DMFT is never
an exact numerical approach, although a DMFT-based approach
for a 3D system could possibly yield better results than 
direct QMC or DMRG calculations.
It should be noticed that in the DMFT framework
the self-energy has to be determined 
self-consistently by solving a quantum impurity problem, which is itself a
difficult strongly correlated problem. This quantum impurity problem
is usually solved numerically using one of the standard methods discussed here
(ED, NRG, QMC or DMRG). Therefore, the DMFT approach and 
its extensions can be viewed as 
an (approximate) way of circumventing the limitations of the
standard methods and extend their applicability to large 3D
systems.

In summary, every numerical method has advantages and weaknesses.  
ED is unbiased but is restricted to small clusters. 
For 1D systems DMRG is usually the best method
while for 3D systems only direct QMC simulations are possible.  
QMC techniques represent also the most successful approach 
for non-critical and non-frustrated two-dimensional systems.
There is currently no satisfactory numerical (or analytical)
method for two-dimensional strongly correlated systems
with frustration or in a critical regime.

\section{Exact diagonalization approach \label{sec:ed}}
As stated above, ED is presently probably the best controlled numerical method
because it allows an approximation-free treatment of coupled
electron-phonon models in the whole parameter range. 
As a precondition we have to work with finite systems 
and apply a well-defined  truncation procedure for the phonon
sector (see subsect.~\ref{subsec:hs_bc}). At least for the 
single-electron Holstein model a  variational basis can 
be constructed in such a way that the ground-state properties 
of the model can be computed numerically exact in the  
thermodynamic limit (cf. subsect.~\ref{subsec:vm}). 
In both cases the resulting numerical problem is to find 
the eigenstates of a (sparse) Hermitian matrix using Lanczos 
or other iterative subspace methods  (subsect.~\ref{subsec:ep}). 
In general the computational
requirements of these eigenvalue algorithms are determined by
matrix-vector multiplications (MVM), which have to be implemented in 
a parallel, fast and memory saving way on modern supercomputers. 
Extensions for the calculation
of dynamical quantities have been developed on the
basis of Lanczos recursion and kernel polynomial expansions
(cf. subsect.~\ref{sec:sp}).
Quite recently  cluster perturbation theory (CPT) has been used
in combination with these techniques to determine the single-particle
electron and phonon spectra. 

\subsection{Many-body Hilbert space and basis construction\label{subsec:hs_bc}}
\subsubsection{Basis symmetrization\label{subsubsec:bs}}
The total Hilbert space of  
Holstein-Hubbard type  models~(\ref{holstein}) can be written as the tensorial
product space of electrons and phonons, spanned by the
complete basis set 
$\left\{|b\rangle=|e\rangle\otimes |p\rangle\right\}$
with 
\begin{equation}
|e\rangle = \prod_{i=1}^N \prod_{\sigma = \uparrow,\downarrow} 
(c_{i\sigma}^{\dagger})^{n_{i\sigma,e}}  
|0\rangle_{e}\quad\mbox{and}\quad
|p\rangle = \prod_{i=1}^N \frac{1}{\sqrt{m_{i,p} !}} 
(b_i^{\dagger})^{m_{i,p}}|0\rangle_{p}.
\label{basis1}
\end{equation}
Here $n_{i\sigma,e} \in \{0,1\}$, i.e. the electronic  
Wannier site $i$ might be empty, singly or doubly occupied, whereas 
we have no such restriction for the phonon number, 
$m_{i,p}\; \in \{0,\ldots,\infty\}.$ Consequently, 
$e=1,\ldots,D_{e}$ and $p=1,\ldots,D_{p}$ label 
basic states of the electronic and phononic subspaces having 
dimensions $D_{e}=\genfrac(){0cm}{1}{N}{N_{e,\sigma}} 
\genfrac(){0cm}{1}{N}{N_{e,-\sigma}} $ 
and $D_{p}=\infty$, respectively. 
Since the Holstein Hubbard Hamiltonian 
commutes with the total electron number operator
$\hat{N}_{e}=
\sum_{i=1}^N(n_{i,\uparrow}+n_{i,\downarrow})$,
$\hat{N}_{e,\sigma}=\sum_{i=1}^Nn_{i,\sigma}$ (we
used the `hat' to discriminate operators from the 
corresponding particle numbers), 
and the $z$-component of the total spin $S^{z}=
\frac{1}{2}\sum_{i=1}^N(n_{i,\uparrow}-n_{i,\downarrow})$, 
the basis $\left\{|b\rangle\right\}$ has been constructed for    
fixed $N_{e}$ and $S^z$.

To further reduce the dimension of the total Hilbert space,
we can exploit the space group symmetries 
[translations ($G_T$) and point group operations ($G_L$)] 
and the spin-flip invariance [($G_S$); $S^z=0$ -- subspace only].
Clearly, working on finite bipartite clusters in 1D or 2D 
(here $N=k^2+l^2$, $k$ and $l$ are both even or odd integers) 
with periodic boundary conditions (PBC),
we do not have all the symmetry properties of the underlying 1D or 
2D (square) lattices~\cite{BWF98}. Restricting ourselves to the 1D
non-equivalent irreducible representations of the group 
$G(\vec{K})=G_T\times G_L(\vec{K})\times G_S$, 
we can use the projection operator 
${\cal P}_{\vec{K},rs}=[g(\vec{K})]^{-1} 
\sum_{{\cal G} \in G(\vec{K})}\chi_{\vec{K},rs}^{(\cal G)}\;{\cal G}$
(with $[H,{\cal P}_{\vec{K},rs}]=0$, 
${\cal P}_{\vec{K},rs}^{\dagger}={\cal P}_{\vec{K},rs}$ and
${\cal P}_{\vec{K},rs}\;{\cal P}_{\vec{K}^{\prime},r^{\prime}s^{\prime}}
={\cal P}_{\vec{K},rs}\;  
\delta_{\vec{K},\vec{K}^{\prime}}\;\delta_{r,r^{\prime}}\;
\delta_{s,s^{\prime}}$) 
in order to generate a new symmetrized basis set:
$\{|b\rangle\} \stackrel{\cal P}{\to}  
\{|\tilde{b}\rangle\}$. ${\cal G}$ denotes the  $g(\vec{K})$ elements of 
the group $G(\vec{K})$ and $\chi_{\vec{K},rs}^{(\cal G)}$ 
is the (complex) character of 
${\cal G}$ in the $[\vec{K},rs]$ representation, where  
$\vec{K}$ refers to one of the $N$ allowed wave vectors in the 
first Brillouin zone,  $r$ labels the irreducible representations 
of the little group of $\vec{K}$, $G_L(\vec{K})$, and $s$
parameterizes $G_S$.   
For an efficient parallel implementation of the MVM 
it is extremely important   
that the symmetrized basis can be constructed preserving the tensor 
product structure of the Hilbert space, i.e., 
\begin{equation}
\{|\tilde{b}\rangle=
N^{[\vec{K}rs]}_{\tilde{b}}\, {\cal P}_{\vec{K},rs}\,
\left[ |\tilde{e}\rangle
\otimes |p\rangle\right]\} ,
\label{basis2} 
\end{equation}
with  $\tilde{e}=1,\ldots, \tilde{D}_{e}^{g(\vec{K})}$ 
$[\tilde{D}_{e}^{g(\vec{K})}\sim D_{e}/g(\vec{K})]$. The
$N^{[\vec{K}rs]}_{\tilde{b}}$ are normalization factors. 

\subsubsection{Phonon Hilbert space truncation\label{subsubsec:hst}}
Since the Hilbert space associated to the phonons is infinite
even for a finite system, we apply a truncation procedure~\cite{WRF96}
retaining only basis states with at most $M$ phonons:  
\begin{equation}
\{ |p\rangle\; ; \;m_p=\sum^N_{i=1} m_{i,p} \le M \}.
\label{eq:cutoff}
\end{equation}
The resulting Hilbert space has a total dimension 
$\tilde{D}=\tilde{D}_{e}^{g(\vec{K})}\times D_{p}^M$ with 
$D_{p}^M =\frac{(M+N)!}{M!N!}$, and a general state 
of the Holstein Hubbard model is represented as   
\begin{equation}
|\psi_{\vec{K},rs}\rangle = 
\sum_{\tilde{e}=1}^{\tilde{D}_{e}^{g(\vec{K})}} 
\sum_{p=1}^{D^M_{p}} c_{\tilde{e}p}\,
|\tilde{b}\rangle. 
\end{equation}
It is worthwhile to point out that, switching from a real-space representation
to a momentum space description, our truncation scheme takes 
into account all dynamical phonon modes. This has to be 
contrasted with the frequently used single-mode 
approach~\cite{AP98}. 
In other words, depending on the model parameters and the band filling, 
the system ``decides'' by itself how the $M$ phonons will be 
distributed among the independent Einstein oscillators related 
to the $N$ Wannier sites or, alternatively, among the $N$ different 
phonon modes in momentum space. Hence with the 
same accuracy phonon dynamical effects on lattice distortions 
being quasi-localized in real space 
(such as polarons, Frenkel excitons, \ldots) or in momentum space 
(like charge-density-waves, \ldots)
can be studied. 

\begin{figure}[t]
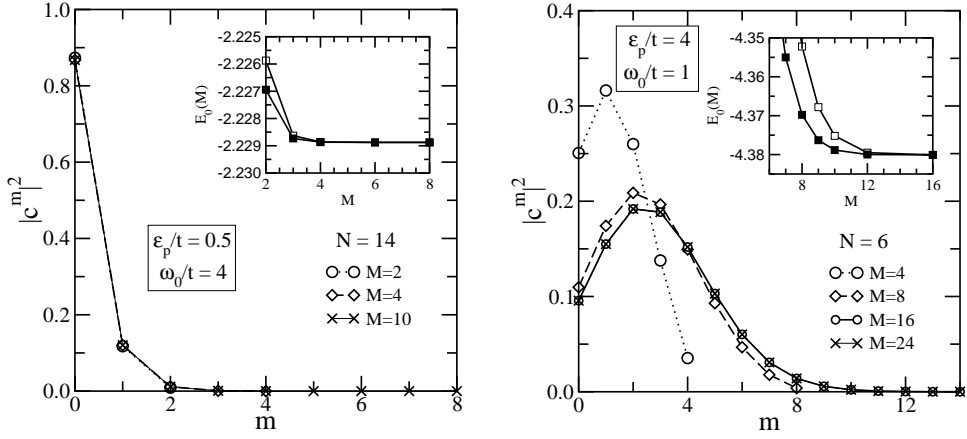

\includegraphics[width=.45\linewidth,clip]
{jeckel_fehske_fig2a.eps}\hspace*{0.5cm}
\includegraphics[width=.45\linewidth,clip]
{jeckel_fehske_fig2b.eps}
\caption{Convergence of the phonon distribution 
function $|c^m|^2(M)$ and ground-state
energy $E_0(M)$ (inset; here filled symbols give the results
obtained separating the $Q=0$ phonon mode) 
as a function of the maximum number of 
phonons $M$ retained. Results are given 
for the Holstein model on a 1D lattice with
$N=14$ (a) and $N=6$ (b) sites (PBC), where the parameters
$\varepsilon_p$ and $\omega_0$ are  the same 
as for the CPT calculation
presented in fig.~3 of the paper by
Fehske, Alvermann, Hohenadler and Wellein.
}
\label{f:phon_dis}
\end{figure}

Of course, one has carefully to check for the convergence of
the above truncation procedure by calculating the ground-state
energy as a function of the cut-off parameter $M$. 
In the numerical work convergence is assumed to be achieved if 
$E_0$ is determined with a relative error
$\Delta E_0^{(M)} = (E_0(M)-E_0(M-1))/E_0(M)\leq 10^{-6}$\,.
In addition we guarantee that the phonon distribution function 
\begin{equation}
|c^{(m)}|^2(M) = \sum^{\tilde{D}^{g(\vec{K})}_{e}}_{\tilde{e}=1} 
\sum^{D^M_{p}}_{\stackrel{{\displaystyle {}_{p=1}}}{\{m_p=m\}}} 
|c_{\tilde{e}p}|^2,
\end{equation}
which gives the different weights of the $m$-phonon states in the 
ground-state $|\psi_0\rangle$, becomes independent of $M$ 
and $|c^{(M)}|^2(M)\leq 10^{-6}$. 
To illustrate the $M$ dependences of the phonon distribution function
and the ground-state energy, we have shown both quantities 
in fig.~\ref{f:phon_dis} for the single-electron Holstein model 
on rather small lattices. 
Figure~\ref{f:phon_dis}  
proves that our truncation procedure is well controlled
even in the strong EP coupling regime, where multi-phonon states become
increasingly important. 

For the Holstein-type models 
the computational requirements can be further reduced.  
Here it is possible to separate the symmetric 
 phonon mode, $B_{0}=\frac{1}{\sqrt{N}}\sum_{i} b_i$, and to
calculate its contribution to $H$ analytically~\cite{SHBWF05}. 
For the sake of simplicity, we restrict ourselves to the 1D spinless 
case  in what follows. 
Using the momentum space representation of the phonon operators, 
the original Holstein Hamiltonian takes the form
\begin{equation}
H = -t \sum_{ij} (c^{\dagger}_{i}c_{j}^{}
 +c^{\dagger}_{j}c_{i}^{}) - \sqrt{\varepsilon_p\omega_0} \;\sum_{j}(
{B^{\dagger}_{-Q_j}}+ {B^{}_{Q_j}}) {n}_{Q_j}^{} +\omega_0
\sum_{j} B^{\dagger}_{Q_j} B^{}_{Q_j}
\end{equation}
with $B_{Q_j}^{\dagger} = {\cal U}_{j,i} b^{\dagger}_i$,  
$B_{Q_j}^{}={\cal U}_{j,i}^{*} b^{}_i ={\cal U}_{-j,i}^{} b^{}_i$, 
and $n_{Q_j}= \sum_{i}{\cal U}_{j,i} n_{i}^{}$,
where ${\cal U}_{j,i}= (1/\sqrt{N})\times$\\
$ \exp{\{i Q_j R_i\}}$ and 
$Q_j$ ($R_i$) denote the allowed 
momentum (translation) vectors of the lattice.
The  $Q=0$ phonon mode couples to 
$n_{0}=N_{e}/\sqrt{N}$ which is a constant 
if working in a subspace with fixed number of electrons.
Thus the Hamiltonian decomposes into
$H=H^\prime+H_{Q=0}$, with 
$H_{Q=0}=- \sqrt{\varepsilon_p\omega_0}\,
({B^{\dagger}_{0}}+{B^{}_{0}})\,n_{0}^{} 
+ \omega_0  B^{\dagger}_{0} B^{}_{0}$. 
Since $[H^\prime  ,  H_{Q=0}]=0$, the eigenspectrum 
of $H$ can be built up by the analytic solution for $H_{Q=0}$ 
and the numerical results for $H^\prime$.
Using the unitary transformation 
\begin{equation}
\label{UTransS}
{\cal \bar{S}} (N_{e}) =  \exp{ \left\{ - \frac{N_{e}}{\sqrt{N}}
\sqrt{\frac{\varepsilon_p}{\omega_0}}
  ({B^{\dagger}_{0}}-{B^{}_{0}}) \right\} }
\end{equation}
which introduces a shift of the phonon operators 
($B^{}_{0} \rightarrow B^{}_{0}+ \frac{N_{e}}{\sqrt{N}}
\sqrt{\frac{\varepsilon_p}{\omega_0}}$), 
we easily find the diagonal form of 
$\bar{H}_{Q=0} = \omega_0  B^{\dagger}_{0} B^{}_{0} 
- \varepsilon_pN^2_{e}/N$.                 
It represents a harmonic oscillator with 
eigenvalues and eigenvectors
$\bar{E}_{\bar{l}} = 
\omega_0 \bar{l} -\varepsilon_p \frac{N^2_{el}}{N}$ and
$| \bar{l} \rangle = \frac{1}{\sqrt{ \bar{l} !} }
  (B^{\dagger}_{0} )^{\bar{l}} | 0 \rangle$. 
The corresponding eigenenergies and  eigenvectors of 
$H_{Q=0}$ are $E_{l}=\bar{E}_{\bar{l}}$ and 
$|l (N_{e} )\rangle =  {\cal \bar{S}}^{\dagger} (N_{e}) | \bar{l} \rangle$, 
respectively. That is, in the eigenstates of the Holstein model
a homogeneous lattice distortion occurs.
Note that the homogeneous lattice distortions are different 
in subspaces with different electron number. Thus excitations due to 
lattice relaxation processes will show up in the one-particle 
spectral function.
Finally, eigenvectors and eigenenergies of $H$ can be constructed by  
combining the above analytical result with the numerically determined 
eigensystem ($E^\prime_{n}$; $|\psi_{n}^{\prime} \rangle$) 
of $H^\prime$:
$E_{n,l}^{\prime}= E_{n}^{\prime} + \omega_0 l 
- \varepsilon_p N^2_{e}/N$ and 
$|\psi_{n,l} \rangle = |\psi_{n}^{\prime} \rangle \otimes |l ( N_{e})
\rangle$.

\subsection{Variational ED method\label{subsec:vm}}
In this section we briefly outline a very efficient variational method 
to address  the (one-electron) Holstein polaron problem numerically
in any dimension. 
The approach was developed by Bon\v{c}a {\it et  al.}~\cite{BTB99,KTB02} 
and is based on a clever way of constructing the EP Hilbert 
space which can be systematically
expanded in order to achieve high-accuracy results with rather modest 
computational resources. 

\begin{figure}[t]
\begin{minipage}{0.45\linewidth}
\includegraphics[width=\linewidth,clip]
{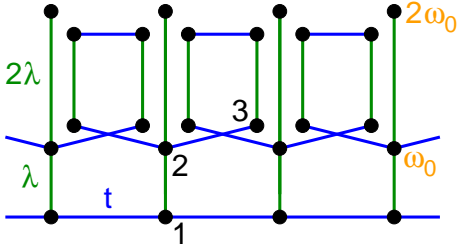}
\end{minipage}\hspace*{0.3cm}
\begin{minipage}{0.50\linewidth}
\caption{Small variational Hilbert space for the 1D polaron problem.
Basis states are represented by dots, off-diagonal matrix elements 
by lines. Vertical bonds create or destroy phonons with frequency
$\omega_0$. Horizontal bonds correspond to electron hops $(\propto t)$.
Accordingly, state $|1\rangle$ describes an electron at the 
origin (0) and no phonon,
state $|2\rangle$ is an electron and one phonon both at site 0,
$|3\rangle$ is an electron at the nearest neighbor-site site 1,
and a phonon at site 0, and so on. The figure is re-drawn from 
ref.~\cite{BTB99}. 
\vspace*{0.2cm}
\label{f:vm}}
\end{minipage}
\end{figure}

The authors built up the variational space starting from an initial state,
e.g. the electron at the origin, and acting repeatedly ($L$ times) with the
off-diagonal diagonal hopping ($t$) and EP coupling ($\lambda$) terms
of the Hamiltonian~(see fig.~\ref{f:vm}). A basis state is added
if it is connected by a non-zero $t$- or $\lambda$-matrix element
to a state previously in the space, i.e., states in generation $l$ are
obtained by acting $l$ times with off-diagonal terms. Only one copy
of each state is retained. Importantly, all translations of these 
states on an infinite lattice are included. According to Bloch's
theorem each eigenstate can be written as $\psi=\mbox{e}^{{\rm i} kj} a_L$,
where $a_L$ is a set of complex amplitudes related
to the states in the unit cell, e.g. $L=7$ for the small variational
space shown in fig.~\ref{f:vm}. For each momentum $K$ the resulting 
numerical problem is then to diagonalize a Hermitian $L\times L$
matrix. While the size of the Hilbert space increases as (D+1)$^L$,
the error in the ground-state
energy decreases exponentially with $L$. Thus in most cases 
$10^4$-$10^6$ basis states are sufficient to obtain an 8-10 digit
accuracy for $E_0$. The ground-state energy calculated this way is 
variational for the infinite system.

\subsection{Solving the eigenvalue problem\label{subsec:ep}}
To determine  the eigenvalues of large sparse Hermitian matrices,
iterative (Krylov) subspace methods like Lanczos~\cite{CW85} and variants of
Davidson~\cite{Da75} diagonalization techniques 
are frequently applied. These algorithms 
contain basically three steps:
\begin{itemize}
\item[(1)] project problem matrix ${\bf A}$ $ \in \mathbb{R}^n$ onto a 
subspace $\bar{\bf A}^k \in \mathbb{V}^k$ $(k\ll n)$ 
\item[(2)] solve the eigenvalue problem in $\mathbb{V}^k$ 
using standard routines
\item[(3)] extend subspace $\mathbb{V}^k \to \mathbb{V}^{k+1}$ by 
a vector $\vec{t}\perp \mathbb{V}^k $ and go back to (2).
\end{itemize}
This way we obtain a sequence of approximative inverses of the
original matrix ${\bf A}$.

\subsubsection{Lanczos diagonalization\label{subsubsec:ld}}
Starting out from an arbitrary (random) initial state
$|\varphi_0\rangle$, 
having finite overlap with the true ground state $|\psi_{0}\rangle$,
the Lanczos algorithm recursively generates a set of orthogonal
states (Lanczos or Krylov vectors):
\begin{equation}
|\varphi_{l+1}\rangle=
{\bf H}^{\tilde{D}}|\varphi_{l}\rangle
-a_l|\varphi_{l}\rangle
-b_l^2|\varphi_{l-1}\rangle,
\label{lr1}
\end{equation} 
where
$a_l=\langle\varphi_{l}|{\bf H}^{\tilde{D}}
|\varphi_{l}\rangle/\langle\varphi_{l}|
\varphi_{l}\rangle,
b_l^2=\langle\varphi_{l}|\varphi_{l}\rangle/\langle\varphi_{l-1}|
\varphi_{l-1}\rangle, b_0^2=0$,
and $|\varphi_{-1}\rangle=0$. 
Obviously, the representation matrix  $[{\bf T}^L]_{l,l^{\prime}}=
\langle \varphi_{l}|{\bf H}^{\tilde{D}}|
\varphi_{l^{\prime}}\rangle$ of ${\bf H}^{\tilde{D}}$ is tridiagonal 
in the $L$-dimensional Hilbert space spanned by the
$\{|\varphi_{l}\rangle\}_{l=0,\ldots,L-1}$ ($L\ll\tilde{D}$):  
\begin{equation}
  \label{tm}
  {\bf T}^{L} = \left(
    \begin{array}{cccccc}
      a_{0} & b_{1} & 0 & 0 & 0 & \cdots\\ b_{1} & a_{1} & b_{2} & 0 & 0
      &\cdots\\ 0 & b_{2} & a_{2} & b_{3} & 0 & \cdots\\ 0 & 0 & b_{3} & a_{3}
      & b_{4} & \cdots\\ \vdots & \vdots & \vdots & \vdots & \vdots & \ddots
    \end{array}
  \right).
\end{equation}
Applying the Lanczos recursion~(\ref{lr1}), the eigenvalues $E_n$ and 
eigenvectors $|\psi_{n}\rangle $ of ${\bf H}^{\tilde{D}} $ 
are approximated by 
\begin{equation}
E_n^L\quad\mbox{ and}\quad  
|\psi^L_{n}\rangle=\sum_{l=0}^{L-1}c_{n,l}^L|\varphi_{l}\rangle ,
\label{cnlL}
\end{equation}
respectively, where the L coefficients $c_{n,l}^L$  are 
the components of the ($n-{\rm th}$) eigenvectors
of ${\bf T}^L$ with eigenvalue $E_n^L$. 
The eigenvalue spectrum of ${\bf T}^L$ can be easily
determined using standard routines from libraries 
such as EISPACK (see http://www.netlib.org).
Increasing $L$ we check for the convergence of an eigenvalue of 
${\bf  T}^L$ in a specific energy range. So we can avoid spurious eigenvalues
for fixed Lanczos dimension $L$ which disappear as one varies $L$~\cite{CW85}.

Note that the convergence of the Lanczos algorithm  is excellent at the
edges of the spectrum (the ground state for example is obtained with
high precession using at most $\sim 100$ Lanczos iterations) 
but rapidly worsens inside the
spectrum. 
So Lanczos is suitably used only to obtain the ground state
and a few low lying excited states.

\subsubsection{Implementation of matrix vector multiplication
  \label{sec:mvi}} 
The core operation of most ED algorithms is a MVM.
It is quite obvious that our matrices are extremely sparse 
because the number of non-zero entries per row of our Hamilton matrix 
scales linearly with the number of electrons.
Therefore a standard implementation of the MVM step uses a sparse storage
format for the matrix, holding the non-zero elements only. 
Two data schemes are in wide use, the compressed row 
storage (CRS) and the jagged diagonal 
storage (JDS) format~\cite{Baea93}, where the
latter is the method of choice for vector computers.
The typical storage requirement per non-zero entry
is 12-16 Byte for both methods, i.e. for a matrix dimension of
$\tilde{D}=10^9$ about one TByte main memory is required 
to store only the matrix elements of the EP Hamiltonian.
Both variants can be applied to any sparse matrix structure 
and the MVM step can be done in parallel by using a 
parallel library such as PETSc
(see http://www-unix.mcs.anl.gov/petsc/petsc-as/).

To extend our EP studies to even larger matrix sizes we 
store no longer the non-zero matrix elements but generate them
in each MVM step. Of course, at that point standard 
libraries are no longer useful and a parallel code 
tailored to each  specific class of Hamiltonians must be developed.
For the Holstein-Hubbard EP model we have established a massively 
parallel program using the Message Passing Interface (MPI) 
standard. The minimal total memory requirement of this
implementation is three vectors with Hilbert space  dimension.

The parallelization approach follows the inherent natural parallelism of
the Hilbert space, which can be constructed as the tensorial product space
of electrons and phonons 
$\{|\tilde{b}\rangle=|\tilde{e}\rangle\otimes|p\rangle\}$  
(cf. subsubsect.~\ref{subsubsec:bs}).
Assuming, that the electronic dimension ($\tilde{D}_{e}$) is a multiple 
of the number of processors used ($N_{\rm cpu}$) we can easily 
distribute the electronic basis states among these processors, 
i.e. processor $i (0 \leq i \leq N_{\rm cpu}-1)$ is holding the 
basis states ($\tilde{e}_i = i \tilde{D}_e/N_{\rm cpu}+1,
\ldots , (i+1) D_e/N_{\rm cpu}$). 
As a consequence of this choice only the electronic hopping term 
generates inter-processor communication in the MVM
while all other (diagonal electronic) contributions can be 
computed locally on each processor. 

Furthermore, the communication pattern remains constant within a single run 
for all MVM steps and the message sizes (at least $D_{p}$ words) 
are large enough to ignore the latency 
problems of modern interconnects.
Using  supercomputers with hundreds of processors and one TBytes 
of main memory, such as IBM p690 clusters or SGI Altix systems, 
we are able to run simulations up to a matrix dimension of 
$30 \times 10^{9}$. 

\subsection{Algorithms for estimating spectral
  functions\label{sec:sp}}
The numerical calculation of spectral functions
\begin{eqnarray}
A^{\cal O}(\omega)&=&-\lim_{\eta\to 0^+}\frac{1}{\pi} \mbox{Im} \left[
\langle\psi_{0}
|{\bf O}^{\dagger}\frac{1}{\omega - {\bf H} +E_0 +i\eta}{\bf O}^{} 
|\psi_{0}\rangle\right]\nonumber\\&=&
\sum_{n=0}^{\tilde{D}-1}|\langle\psi_{n}|{\bf O}|
\psi_{0}\rangle |^{2}\delta [\omega - (E_{n} - E_{0})],
\label{specfu}
\end{eqnarray}
where ${\bf O}$ is the matrix representation of a certain 
operator ${\cal O}$ (e.g., the creation operator
$c_{k}^\dagger$ of an electron with wave number $k$ if one wants 
to calculate the single-particle spectral function; or the current operator
$\hat{\jmath} = - \mbox{i} e t\sum_{i}(c_{i}^{\dagger} 
  c_{i+1}^{} - c_{i+1}^{\dagger} 
  c_{i}^{})$ if one is interested in the optical conductivity),  
involves the resolvent of the Hamilton 
matrix ${\bf H}$. Once we have obtained the 
eigenvalues and eigenvectors of $H$ we can plug them into
eq.~(\ref{specfu}) and obtain directly the corresponding  
dynamical correlation or Green functions.
In practice this `naive' approach is applicable  for small
Hilbert spaces only, where the complete 
diagonalization of the Hamilton matrix is feasible.

For the typical EP problems under investigation    
we deal with Hilbert spaces having total dimensions $\tilde{D}$ 
of $10^6$-$10^{11}$. Finding all eigenvectors and eigenstates 
of such  huge Hamilton matrices is impossible, 
because the CPU time required for
exact diagonalization of ${\bf H}$ scales as 
$\tilde{D}^3$ and memory as $\tilde{D}^2$. 
Fortunately, there exist very accurate and well-conditioned 
linear scaling algorithms for a direct approximate calculation of 
$A^{\cal O}(\omega)$. 

\subsubsection{Lanczos recursion method\label{subsubsec:sdm}}
Having determined the ground state $|\psi_{0}^L\rangle$ by the 
Lanczos technique, we can use again the recursion relation~(\ref{lr1}),
but with the initial state
$|\varphi_0\rangle={\bf O}|\psi^L_{0}\rangle/\sqrt{
\langle \psi^L_{0}|{\bf O}^{\dagger}{\bf O}|\psi^L_{0}
\rangle}$,
to determine within the so-called {\it Lanczos recursion method} (LRM)
or {\it spectral decoding method} (SDM) an approximative spectral function,
\begin{equation}
\bar{A}^{\cal O}(\omega)=\sum_{n=0}^{L-1} |c_{n,0}^L|^2 
\langle\psi_{0}|{\bf O}^{\dagger}{\bf O}|\psi_{0}\rangle\,
\delta[\omega -(E_n^L-E_0^L)], 
\label{asdm}
\end{equation} 
or equivalently
\begin{equation}
  \label{equ:lrmform}
  \bar{A}^{\cal O}(\omega) = -\lim_{\eta\to 0^+}\frac{1}{\pi}\mbox{Im} {\,\,
      \frac{ \langle\psi_{0}| {\bf O}^{\dagger} {\bf O} |\psi_{0}\rangle}
      {\omega+\mbox{i}\eta -a_{0}-\cfrac{b_{1}^{2}} {\omega+{\rm
      i}\eta -a_{1}-\cfrac{b_{2}^{2}} {z-a_{2}-\cdots}}}
    }\,,
\end{equation}
which is built up by $L$ $\delta$-peaks.

Of course,
the true spectral function $A^{\cal O}(\omega)$ has  $\tilde{D}$
$\delta$-peaks. According to the  Lanczos phenomenon, the approximated 
spectral weights and positions of the peaks converge to their true values 
with increasing $L$. Some of the main problems of the LRM/SDM are:
(i) The convergence is not uniform in the whole energy range.  
(ii) There exist so-called spurious peaks, which appear 
and disappear as $L$ is increased, i.e., when the iteration proceeds.
(iii) Without computationally expensive re-orthogonalization only a 
few hundred iterations are possible. 

\subsubsection{Kernel polynomial method\label{subsubsec:kpm}}
The idea behind a conceptionally different approach, the
{\it kernel polynomial method} (KPM) (for a review see~\cite{WWAF05}), is 
to expand  $A^{\cal O}(\omega)$ 
in a finite series of $L+1$ Chebyshev polynomials $T_m(x)= \cos [m
\arccos(x)]$. Since the Chebyshev polynomials are defined 
on the real interval $[-1,1]$, we apply first a simple
linear transformation to the Hamiltonian and all energy scales:
${\bf X}=({\bf H}-b)/a$, $x=(\omega -b)/a$, 
$a=(E_{\rm max}-E_{\rm min})/2(1-\epsilon)$,
and $b=(E_{\rm max}+E_{\rm min})/2$  
(the small constant $\epsilon$ is introduced in order to avoid
convergence problems at the endpoints of the interval 
 --a typical choice is $\epsilon \sim 0.01$ which has only
1\% impact on the energy resolution~\cite{SR97}). 
Then the expansion reads
\begin{equation}
A^{\cal O}(x)=\frac{1}{\pi
\sqrt{1-x^{2}}}\left(\mu_{0}^{\cal O}+
2\sum_{m=1}^{L}\mu_{m}^{\cal O}T_{m}(x)\right),
\label{akpm}
\end{equation}
with the coefficients (moments) 
\begin{equation}
\mu_m^{\cal O}=\int_{-1}^{1}dx\,T_{m}(x)A^{\cal O}(x)=\langle
\psi_{0}| {\bf O}^{\dagger}T_{m}({\bf X}){\bf O}^{}|\psi_{0}\rangle.
\label{mkpm}
\end{equation}
Equation~(\ref{akpm}) converges to the correct function for $L\to\infty$.
Again the moments 
\begin{equation}
\mu_{2m}^{\cal O}=2\langle\phi_m|\phi_m\rangle -\mu_0^{\cal O}
\quad\mbox{and}\quad
\mu_{2m+1}^{\cal O}=2\langle\phi_{m+1}|\phi_m\rangle -\mu_1^{\cal O}
\label{moments2}
\end{equation}
can be efficiently obtained by repeated parallelized MVM, 
where $|\phi_{m+1}\rangle
=2{\bf X}|\phi_m\rangle -|\phi_{m-1}\rangle$
but now $|\phi_1\rangle={\bf X}|\phi_0\rangle$ and  
$|\phi_0\rangle={\bf O}|\psi_0\rangle$ 
with $|\psi_0\rangle$ 
determined by Lanczos ED.

As is well known from Fourier expansion, 
the series~(\ref{akpm}) with $L$ finite 
suffers from rapid oscillations (Gibbs phenomenon)
leading to a poor approximation to $A^{\cal O}(\omega)$.
To improve the approximation 
the moments $\mu_n$ are modified $\mu_n \to g_n \mu_n$, 
where the damping factors $g_n$ 
are chosen to give the `best' approximation for a given $L$.
This modification is equivalent to a convolution of the infinite series
with a smooth approximation $K_L(x,y)$ to $\delta(x-y)$,
a so-called approximation kernel.
The appropriate choice of this kernel, that is of $g_n$, 
e.g. to guarantee positivity of $A^{\cal O}(\omega)$, 
lies at the heart of KPM. 
We mainly use the Jackson kernel 
which results in a uniform approximation whose resolution 
increases as $1/L$, but for the determination of the 
single-particle Green functions below we use a Lorentz kernel 
which mimics a finite imaginary part $\eta$ 
in eq.~(\ref{specfu}), see~\cite{WWAF05}.

In view of the uniform convergence of the expansion,  
KPM is a method tailored to the calculation of spectral properties.
Most important, spectral functions obtained via KPM are not subject to
uncontrolled or biased approximations:
The accuracy of its outcome depends only on the expansion depth
$L$, and can be made as good as required by just increasing $L$.
Of course one is restricted to finite systems of moderate size
whose associated Hamilton matrix does not exceed available
computational resources.

\subsubsection{Cluster perturbation theory (CPT)\label{subsubsec:cpt}}

The spectrum of a finite system of $N$ sites which we obtain through
KPM differs in many respects from that in the thermodynamic limit
$N\to\infty$,
especially it is obtained for a finite number of momenta 
$K=\pi\, m/N$ only.
The most obvious feature is the resulting discreteness of energy levels
which is a property already of the non-interacting system.
While we cannot easily increase $N$ without going beyond
computationally accessible Hilbert spaces,
we can try to extrapolate from a finite to the infinite system. 

For this purpose we first calculate the Green function $G^c_{ij}(\omega)$
for all sites $i,j=1,\dots,N$ of a $N$-size cluster with open
boundary conditions, and then recover the infinite lattice
by pasting identical copies of this cluster at their edges.
The `glue' is the hopping $V$
 between these clusters, where  $V_{kl}=t$ for $|k-l|=1$ and
$k,l \equiv 0,1 \bmod N$, 
which is dealt with in first order perturbation theory.
Then the Green function $G_{ij}(\omega)$ of the infinite lattice
is given through a Dyson equation
\begin{equation}
  G_{ij}(\omega) = G^c_{ij}(\omega) + \sum_{kl} G^c_{ik}(\omega) V_{kl} 
  G_{lj}(\omega),
\end{equation}
where indices of $G^c(\omega)$ are counted modulo $N$.
Obviously this order of perturbation in $V$ is exact for the
non-interacting system. We thus get rid of the discreteness addressed above.
The Dyson equation is solved by Fourier transformation over momenta
$K= k N$ corresponding to translations by $N$ sites
\begin{equation}
  G_{ij}(K,\omega) 
= \left[ \frac{G^c(\omega)}{1-V(K)G^c(\omega)} \right]_{ij} ,
\end{equation}
from which one finally obtains
\begin{equation}
G(k,\omega) = \frac{1}{N} \sum_{i,j=1}^{N} G^c_{ij}(N k,\omega) 
\exp(- \mathrm{i} k (i-j) ) .
\end{equation}

In this way, which is called  CPT~\cite{SPP00}, 
we obtain a Green function $G(k,\omega)$ with continuous
momenta $k$ from the Green function $G^c_{ij}(\omega)$ on a finite cluster.
Two approximations are made, 
one by using first order perturbation theory in $V=t$,
the second on assuming translational symmetry in $G_{ij}(\omega)$
which is only approximatively satisfied.

In principle, the CPT spectral function $G(k,\omega)$ does not 
contain any more
information than the cluster Green function $G^c_{ij}(\omega)$  
 already does.
But extrapolating to the infinite system it gives a first hint at the
scenario in the thermodynamic limit. 
However, CPT does not describe effects
which only occur on large length scales,
like Anderson localization
(see the paper by Fehske, Bronold and Alvermann)
or the critical behavior at a phase transition.
Providing direct access to spectral functions, still without relying
on possibly erroneous approximations, 
CPT occupies a niche
between variational approaches like (D)DMRG 
(see sect.~\ref{sec:dmrg} and~\ref{sec:dynamical})
and methods directly working in the thermodynamic limit
like the variational ED method~\cite{BTB99}.

\section{Density matrix renormalization group approach \label{sec:dmrg}}

The Density Matrix Renormalization Group (DMRG)
is one of the most powerful 
numerical techniques for studying many-body systems.
It was developed by Steve White~\cite{Steve_dmrg} in 1992 to overcome
the problems arising in the application of the standard Numerical 
Renormalization Group (NRG) 
to quantum lattice many-body systems
such as the Hubbard model~(\ref{hubbard}). 
Since then the approach has been extended to a great variety of 
problems,
from Classical Statistical Physics
to Quantum Chemistry
(including {\it ab initio} calculations of electronic 
structures in molecules) and, recently, to
Nuclear Physics and the Physics
of Elementary Particles and Fields.
A review article on DMRG and its numerous applications
has recently been published~\cite{Uli_review}. Additional
information
can also be found on the DMRG web page at {\it http://www.dmrg.info}.
A detailed discussion of the basic DMRG algorithms and their
implementation has been published in ref.~\cite{Reinhard_review}.
Readers interested in a simple example 
should consider the application of DMRG to 
single-particle problems, which is also discussed there.
The source code of a single-particle DMRG program
(in the programming language C$^{++}$)
is now part of the ALPS distribution~\cite{ALPS}.

DMRG techniques for strongly correlated systems
have been substantially improved and extended since their
conception and
have proved to be both extremely accurate for low-dimensional
problems and widely applicable.
They enable numerically exact calculations 
(as good as exact diagonalizations) 
of low-energy properties
on large lattices with up to a few thousand particles and sites
(compared to less than a few tens for exact diagonalizations). 
The calculation of high-energy excitations and dynamical
spectra for large systems has proved to be more difficult 
and will be discussed in sect.~\ref{sec:dynamical}.

\subsection{Renormalization group and density matrix}

Consider a quantum lattice system with $N$ sites
(in general, we are interested in the case $N \rightarrow \infty$
or at least $N \gg 1$).
The Hilbert space of this system is the Fock space of all
properly symmetrized many-particle wave functions.
As seen in sect.~\ref{sec:ed},
its dimension $D$ grows exponentially with the number of sites
$N$.
Obviously, a lot of these states are not necessary to investigate
specific properties of a model such as the ground state.
Thus, a number of methods have been developed to perform  
a projection onto a subspace of 
dimension $d \ll D$ and then an exact diagonalization of the 
Hamiltonian in this subspace.

\begin{figure}[t]
\begin{center}
\includegraphics[width=2.5cm]{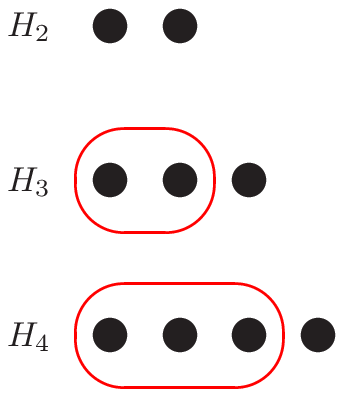}
\includegraphics[width=4.0cm]{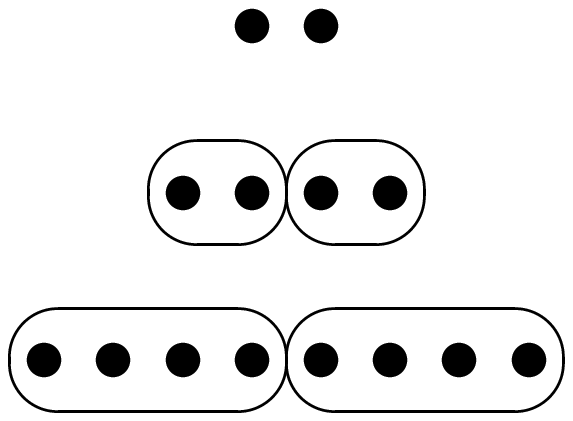}
\includegraphics[width=5.4cm]{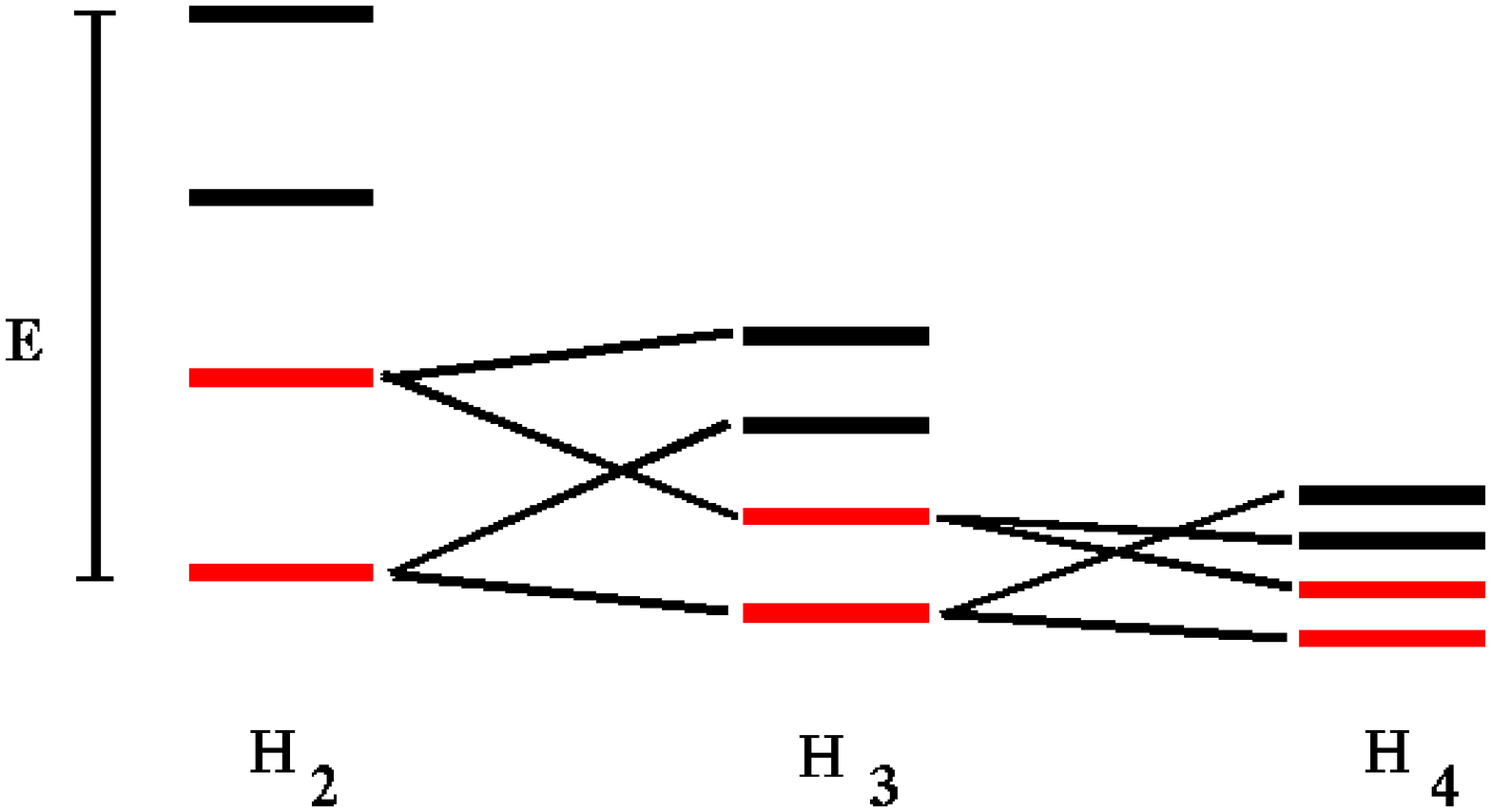}
\caption{Schematic representations of the usual
NRG algorithm (left), an alternative NRG algorithm 
(middle) and the energy levels
in the effective Hamiltonian $H_{\ell}$ for increasing 
subsystem system size $\ell$ (right).}
\label{fig:nrg}
\end{center}
\end{figure}

Such an approach, called the Numerical Renormalization Group (NRG),
was developed by Wilson 30 years ago to solve the Kondo impurity 
problem~\cite{Wilson}.
The key idea is the decomposition of the system into
subsystems with increasing size 
(see the paper on the NRG method by Hewson).
The subsystem size is increased by one site at each step
as shown in fig.~\ref{fig:nrg}.
Each subsystem is diagonalized successively and the information
obtained is used to truncate the Hilbert
space before proceeding to the next larger subsystem.
Let $m$ and $n$ be the dimension of the (effective)
Hilbert spaces associated 
with the subsystem made of the first $\ell$ sites
and with the site $\ell+1$, respectively.   
A basis of dimension $d=mn$ for the next subsystem with $\ell+1$ sites
is built as a tensor product of the subsystem and site bases. 
Assuming that $d$ is small enough, the effective Hamiltonian
can be fully diagonalized in the tensor-product basis.
The energy is used as a criterion to truncate the subsystem Hilbert
space.  
The lowest $m$ eigenstates are kept to form a new basis
while the high-energy eigenstates are discarded. 
The new subsystem with $\ell+1$ sites and an effective Hilbert space 
of dimension $m$
can be used to start the next iteration.
In summary, this procedure provides a transformation
$H_{\ell +1} = R[  H_{\ell}]$
forming effective Hamiltonians of fixed dimension $m$ which
describe the low-energy physics of increasingly larger systems.
In this transformation the high-energy states are steadily traced
out as the system grows as illustrated in fig.~\ref{fig:nrg}.
Such a transformation $R$ is usually called a renormalization group (RG)
transformation.
A different implementation of the NRG idea is possible
if the system is homogeneous like in the Hubbard model.
A copy of the current subsystem can be substituted 
for the added site in the above procedure. 
Thus the subsystem size doubles at every iteration
as illustrated in fig.~\ref{fig:nrg}.

\begin{figure}[t]
\begin{center}
\includegraphics[width=5.5cm]{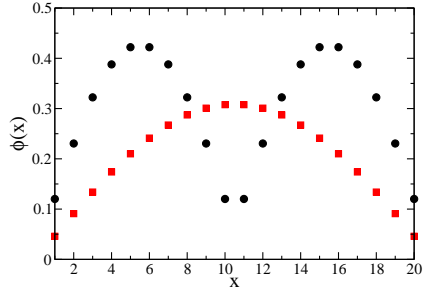}
\caption{Ground-state wavefunctions of the tight-binding
particle-in-the-box problem for two systems of $N=$10 sites
(circles) and one system of $N=20$ sites (squares).
}
\label{fig:pib}
\end{center}
\end{figure}

The NRG method is very accurate for the Kondo problem 
and more generally for quantum impurity problems.
Unfortunately, NRG and related truncation schemes have 
proved to be unreliable for quantum lattice systems such
as the Hubbard model~\cite{nrg_hub}.
It is easy to understand the failure of the standard NRG in those cases.
A subsystem always has an artificial
boundary at which the low-energy eigenstates of a quantum
lattice Hamiltonian tend to vanish smoothly.
Thus the truncation procedure based on the effective eigenenergies may keep
only states that vanish at the artificial boundary.
As a consequence, at later RG iterations the eigenstates of the
effective Hamiltonian of larger subsystems 
may have unwanted features like nodes where
the artificial boundary of the previous subsystems were located.  
The application of the second NRG algorithm
(in the middle in fig.~\ref{fig:nrg}) to the problem of a quantum particle
in a one-dimensional box gives a clear illustration of this
effect~\cite{Reinhard_review,particle_box}.
The ground state wavefunction for a $N$-site lattice 
$\phi(x) = \sqrt{\frac{2}{N+1}}\sin \left (\frac{\pi x}{N+1} \right )$         
has a minimum (node) where the ground state wavefunction
of the twice larger system ($N \rightarrow 2N$) has a maximum
as seen in fig.~\ref{fig:pib}.
Therefore, 
the low-energy eigenstates of a small system
are not necessarily the best states to form the low-energy
eigenstates of a larger system.
An approach proposed by White and Noack~\cite{particle_box} 
to solve this problem is the construction of
an effective Hamiltonian including the effects 
of the subsystem environment  
to eliminate the artificial boundary.
DMRG is the extension of this idea to interacting many-particle 
problems.

Consider a quantum system which can be split into two parts
called the subsystem and its environment.
A basis of the system Hilbert space 
is given by the tensor product
\begin{equation}
\label{basis}
| i,j \rangle =  | i \rangle \otimes | j\rangle_{\rm E} 
\end{equation}
of basis states $|i\rangle \ (i=1,\dots,d)$ and 
$|j\rangle_{\rm E} \ (j=1,\dots, d_{\rm E})$ for the subsystem
and its environment, respectively.  
The dimension of this basis is $d_{\rm S} = d \cdot d_{\rm E}$.
Any state $|\psi\rangle$ of the system can be expanded in
this basis
$| \psi \rangle =  \sum_{i,j} \psi_{ij} \ | i \rangle \otimes 
| j\rangle_{\rm E}$  .
The most important states in the subsystem to represent
the state $|\psi\rangle$ are given by its reduced density matrix, 
which is obtained by tracing out the states of the environment
\begin{equation}
\label{rho}
\rho_{i, i^{\prime}} = \sum_j \psi^*_{ij} \psi_{i^{\prime}j} .
\end{equation}
This density matrix is symmetric and has $d$ positive eigenvalues
$w_{\alpha} \geq 0$ satisfying the normalization
$\sum_{\alpha} w_{\alpha} =1$.
A new basis of the subsystem,
$| v_{\alpha} \rangle = \sum_i v_{\alpha i} |i\rangle$, 
can be defined 
using the eigenvectors of the density matrix
$ \sum_{i^{\prime}}  \rho_{ i, i^{\prime}} v_{\alpha i^{\prime}}
= w_{\alpha} v_{\alpha i} ;   \alpha,i = 1, \dots, d .
$              
In the new basis the state $| \psi \rangle$ can be written
\begin{equation}
\label{exactWF}
| \psi \rangle = \sum_{\alpha} \lambda_{\alpha} | v_{\alpha} \rangle
\otimes | u_{\alpha} \rangle_{\rm E} ,
\end{equation}
with $\lambda^2_{\alpha} = w_{\alpha} > 0$ and 
normalized states 
$ | u_{\alpha} \rangle_{\rm E} = \frac{1}{\lambda_{\alpha}}
\sum_{i,j} v^*_{\alpha i} \psi_{ij} | j\rangle_{\rm E}  .
$                
Therefore, $w_{\alpha}$ is the probability that a subsystem is
in a state $| v_{\alpha} \rangle$ if the superblock 
is in the state $|\psi \rangle$. 
The density matrix provides an optimal choice for selecting 
the $m$ states of the subsystem to be kept 
in a RG transformation:
keep density matrix eigenstates $| v_{\alpha} \rangle$ with
the largest weights $w_{\alpha}$.
This is the key idea of the density matrix renormalization group
approach.

\subsection{DMRG algorithms}

In a DMRG calculation one first 
forms a new subsystem with $\ell+1$ sites 
and an effective Hilbert space of dimension $d=mn$
by adding a site to the current subsystem with $\ell$ sites
as in the NRG method.
Then one considers a larger system,
called a superblock, which is made of the new subsystem and 
an environment.
A basis of the superblock Hilbert space 
is given by the tensor product~(\ref{basis}).
Assuming initially that we want to compute the system 
ground state only, we then calculate the ground state
$|\psi\rangle$ of the superblock Hamiltonian using
the techniques discussed in subsect.~\ref{subsec:ep}.
Then the reduced density matrix~(\ref{rho}) for the subsystem  
is obtained by tracing out the environment states.
The $m$ density matrix eigenstates $| v_{\alpha} \rangle$ with
the largest weights $w_{\alpha}$ are kept to build an optimal effective
basis of dimension $m$ for the subsystem with $\ell+1$ sites.
This subsystem is then used to start the next iteration and build the
next larger subsystem. 
This procedure defines a generic density matrix RG 
transformation.
Clearly, the accuracy of a DMRG calculation will depend on the
quality of the environment used.
The environment should make up as much as possible of 
the lattice which is not already included in the subsystem.
Constructing such a large and accurate environment is 
as difficult as the original quantum many-body problem.
Therefore, the environment must also be constructed self-consistently
using a density matrix RG transformation.    

\begin{figure}[b]
\begin{center}
\includegraphics[width=5.2cm]{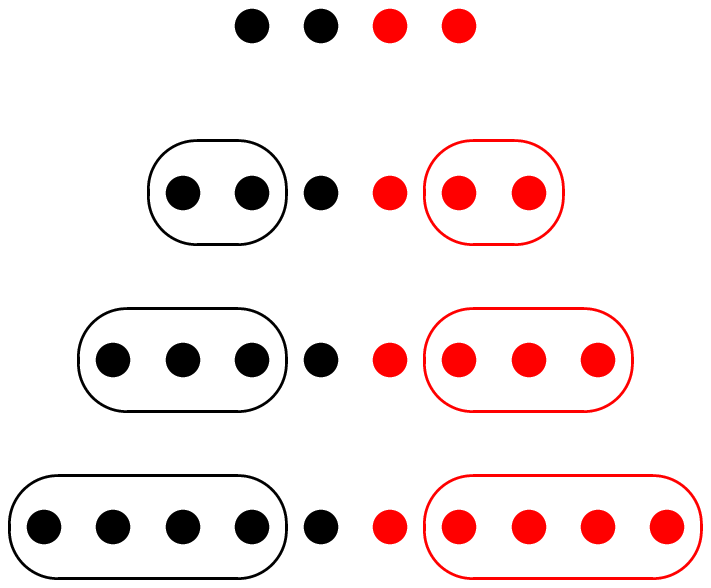}
\includegraphics[width=2.7cm]{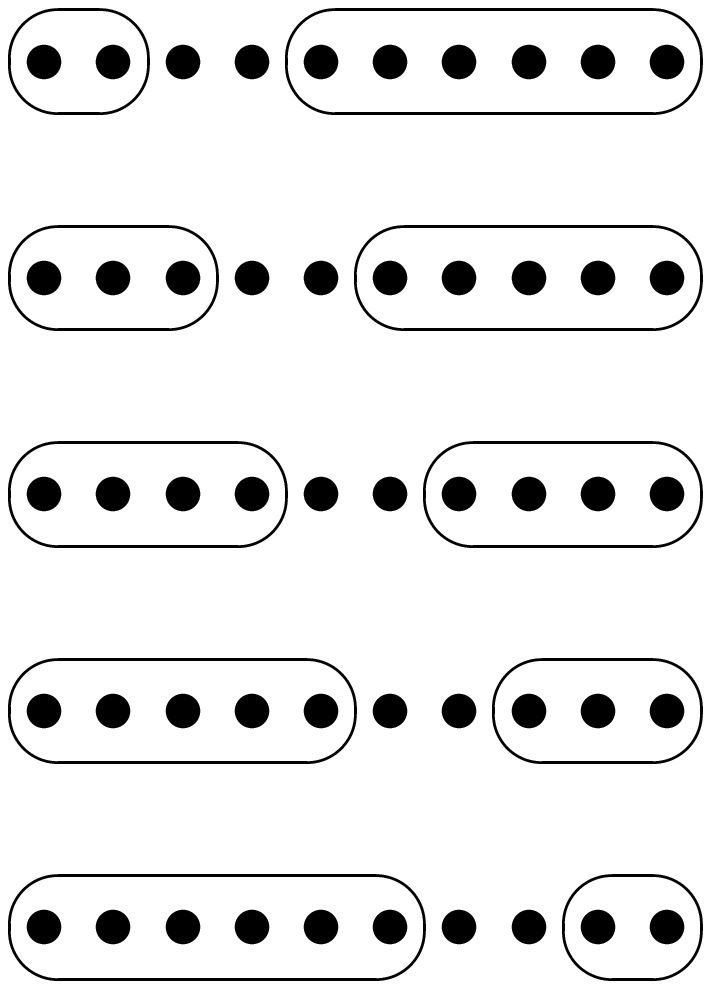}
\caption{Schematic representation of DMRG algorithms. 
Left: Infinite-system algorithm (from top to bottom).
Right: Finite-system DMRG algorithm for a ten-site lattice.
The environment block is on the right-hand side when
going from top to bottom and on the left-hand-side 
when going from bottom to top.}
\label{fig:dmrg}
\end{center}
\end{figure}

In his initial papers~\cite{Steve_dmrg}, 
White described two DMRG algorithms:
the infinite-system method  and the finite-system method.
The infinite-system method is certainly the simplest DMRG algorithm
and is the starting point of many other algorithms.
Its defining characteristic is that the environment
is constructed using a ``reflection'' of the
current subsystem.   
The superblock size increases by two sites at each step as
illustrated in fig.~\ref{fig:dmrg}.   
The system is assumed to be homogeneous and ``symmetric''
to allow this operation.  
Iterations are continued until an accurate approximation of 
an infinite system is obtained.  
The infinite-system method is sometimes used to calculate 
properties of finite systems. 
While this approach may work just fine in many cases, it should be
kept in mind that there is no guarantee of convergence to the
eigenstates of the finite system. 

The finite-system method is
the most versatile and reliable DMRG algorithm
and with some later improvements~\cite{Uli_review,Reinhard_review} it has 
also become a very efficient method.
It is designed to calculate the properties of a finite system
accurately.
The environment is chosen so that the superblock represents
the full lattice at every iteration.
The environments can also be considered as being subsystems and 
are calculated self-consistently using DMRG with the usual
subsystems playing the part of their environment.
Iterations are continued back and forth through every configuration 
of the superblock (i.e., every splitting of the lattice in 
one subsystem and its environment) until convergence.  
This procedure is illustrated in fig.~\ref{fig:dmrg}.
This ensures the self-consistent optimization of the subsystems
and their environments (for a given number $m$ of states kept)
and thus considerably improves the results reliability 
compared to the infinite-system method.

Most DMRG algorithms use only two blocks (the subsystem and its
environment) to form the superblock, but it is possible and useful
for some problems to consider a more complicated configuration.
For instance, one can use four blocks to treat one-dimensional
problems with periodic boundary conditions and 
using several blocks can also be advantageous for systems with
boson degrees of freedom such as phonons~\cite{Robert}.

The DMRG sites usually correspond to
the physical sites of the lattice model investigated, such as
spin, fermionic, or bosonic degree of freedom.
[For a phonon mode (boson), which has an infinite Hilbert space,
a DMRG site represents a finite dimensional basis of the phonon 
states~\cite{Samuel} as already discussed for exact diagonalization methods
(subsect.~\ref{subsec:hs_bc}).]
However, the DMRG sites can also represent a combination of
several physical sites [for instance, the electron and phonon
at each site of the Holstein-Hubbard model~(\ref{holstein})], 
a fraction of
the Hilbert space associated with a given physical site
(as in the pseudo-site method for bosonic degrees of freedom
presented in subsubsect.~\ref{subsubsec:pseudo}).

It is possible to compute several quantum states simultaneously with
DMRG.
In that case, the density matrix is formed as the sum
of the density matrices~(\ref{rho}) 
calculated for each target state.
A target state can be any quantum state which is well-defined 
(and can be computed) in every superblock of a DMRG calculation. 
This feature turns out to be very important
for the calculations of dynamical properties 
(see sect.~\ref{sec:dynamical}).

\subsection{Truncation errors\label{subsec:truncation}}

With DMRG 
an error is made when projecting operators
onto the subspace spanned by the most important $m$
density matrix eigenstates.
This is called the truncation error.
Probably the most important characteristic of a DMRG calculation
is the rate at which the truncation error
decreases with an increasing number $m$ of states kept.
In the most favorable cases (gapped one-dimensional systems
with short-range interactions only and open boundary conditions), 
the accuracy increases roughly exponentially with $m$.
For instance, the ground-state energy of the spin-one Heisenberg chain
on lattices with hundreds of sites can be calculated to an accuracy of
the order of $10^{-10}$ with a modest computational effort.
In very difficult cases (long-range off-diagonal interactions,
two-dimensional systems with periodic boundary 
conditions), the truncation error in the ground state 
energy
can decrease as slowly as $m^{-2}$.
It is possible to calculate exactly the density matrix spectrum
of several integrable models~\cite{Peschel}.
This analysis shows that the distribution of density matrix
eigenvalues $w_{\alpha}$ varies greatly. 
As a result, truncation errors
may fall exponentially with increasing $m$ in some favorable
cases but decrease extremely slowly for other ones.
Correspondingly, DMRG accuracy and performance depend
substantially on the specific problem investigated.

For any target state $| \psi \rangle$ 
written down in its representation~(\ref{exactWF}), 
the truncation of the density matrix basis
corresponds to making an approximation
\begin{equation}
\label{approxWF}
| \psi^{\prime} \rangle = 
\sum_{m \; \mbox{\scriptsize kept states}} 
\lambda_{\alpha} \ | v_{\alpha} \rangle
\otimes | u_{\alpha} \rangle_{\rm E} ,
\end{equation}
which minimizes the error
\begin{equation}
D_m =  \left | |\psi\rangle - |\psi^{\prime} \rangle \right |^2
= \sum_{d-m \; \mbox{\scriptsize discarded states}} 
w_{\alpha} 
= 1 -  
\sum_{m \; \mbox{\scriptsize kept states}}  
w_{\alpha} ,
\end{equation}
for a fixed number $m$ of states kept.
The total weight of the discarded density matrix eigenstates
(discarded weight) $D_m$ is 
related to the truncation errors of physical quantities.
For $D_m \ll 1$ it can be shown that
the truncation error in the eigenenergy of any target
eigenstates of the Hamiltonian $H$ scales linearly with
the discarded weight,  
$                        
E^{\prime}_m - E_{\rm exact} =  c D_m + \mathcal{O}(D_m^2)  \; ,
$                   
where 
$E^{\prime}_m = \langle \psi^{\prime} | H | \psi^{\prime} \rangle/
\langle \psi^{\prime} | \psi^{\prime} \rangle$ is the energy
in the approximate state $ | \psi^{\prime} \rangle$ 
and $c$ is a constant.
For the expectation values of other operators the truncation error
theoretically scales as $\sqrt{D_m}$.  
The discarded weight $D_m$ decreases (and thus the 
accuracy of DMRG results increases) 
when the number $m$ of density matrix eigenstates kept is increased. 
In particular, for large enough $m$ 
the discarded weights vanish at every RG iteration for both 
subsystems and environments and
truncation errors become negligible. 
Therefore, DMRG is an exact numerical method as defined in the
introduction (sect.~\ref{sec:intro}). 
Moreover, if the number $m$ of density matrix eigenstates
kept is so large that the discarded weight is exactly 
zero at every RG iteration,
DMRG becomes equivalent to an exact diagonalization~(sect.~\ref{sec:ed}).

\begin{figure}[t]
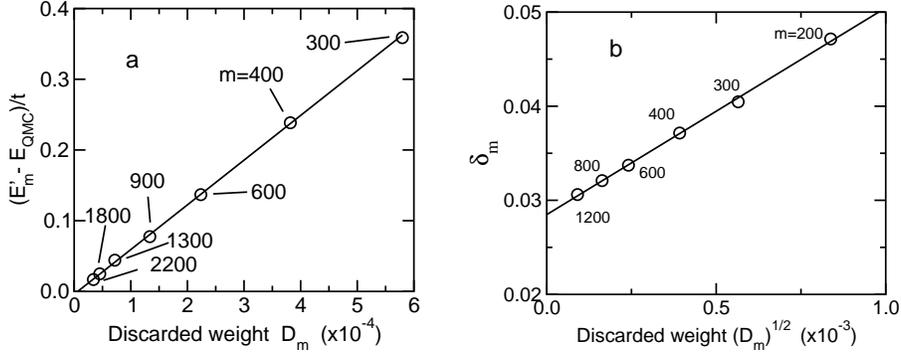

\begin{center}
\includegraphics[width=6.0cm]{jeckel_fehske_fig7a.eps}
\includegraphics[width=6.0cm]{jeckel_fehske_fig7b.eps}
\caption{(a) Ground state energy $E^{\prime}_m$ calculated with DMRG
for a $12\times3$ Hubbard ladder with $U=t$ and
6 holes as a function of the discarded weight
$D_m$ for several number of density matrix 
eigenstates kept $m=300$ to 2200. 
The zero of the energy is given by a (nearly) exact
quantum Monte Carlo result~\cite{Bonca}.
(b) Staggered bond order parameter $\delta_m$ calculated with DMRG
in an extended 1D half-filled Hubbard model~\cite{Eric}
for $U=3t$ and $V=1.5t$ on a 1024-site lattice
as a function of 
$\sqrt{D_m}$ for $m=200$ to 1200. Solid lines are linear fits.
}
\label{fig:truncation}
\end{center}
\end{figure}

In real DMRG applications, series of density matrix basis
truncations are performed in successive superblock configurations. 
Therefore, the measured discarded weights $D_m$
do not represent the real error in the wave function of the
target.
Nevertheless, in most cases, 
the energy truncation error scales linearly
with the average measured discarded weight $D_m$ for $D_m \ll 1$
as shown in fig.~\ref{fig:truncation}(a).
This can (and should) be used to make an $D_m \rightarrow 0$ 
extrapolation of the energy.
For other physical
quantities such as an order parameter,
truncation errors sometimes scale as $(D_m)^r$, with
$r \approx 0.5$ as seen in fig.~\ref{fig:truncation}(b).
This can also be used to estimate truncation errors.  
The measured discarded weight $D_m$ alone should not be used as 
an indication of a calculation accuracy, because truncation
errors for physical quantities can be several orders 
of magnitude larger than $D_m$. 

It should be noted that 
DMRG can be considered as a variational approach.
The system energy 
$E(\psi)= \langle \psi |H|\psi \rangle/\langle \psi |\psi \rangle$
is minimized in a variational subspace of the system
Hilbert space to find the ground-state wavefunction $|\psi_0\rangle$
and energy $E_0 = E(\psi_0)$.
If the ground-state wavefunction is calculated with an error of
the order of $\epsilon \sim \sqrt{D_m}
\ll 1$ (i.e., $|\psi\rangle = |\psi_0\rangle +
\epsilon |\phi\rangle$, with $\langle \phi | \phi \rangle = 1$),
the energy obtained is an upper bound to the exact result and the error
in the energy is of the order of $\epsilon^2 \  (\sim D_m$) 
as in all variational approaches.
For specific algorithms the variational wave function can be written 
down explicitly as a matrix product wave function 
or as a product of local tensors~\cite{Uli_review}. 

The computational effort of a DMRG calculation
usually increases as a power law for increasing system size $N$,
number $m$ of states kept, or DMRG site dimension $n$.
In the most favorable case (1D system with short-range interactions 
only), the CPU time
increases theoretically as $N m^3 n^3$, while the 
amount of stored data is of the order of
$\sim m^2 n^2$.
In most cases, however, $m$ has to be increased with 
the system size $N$ to keep truncation errors constant.

\subsection{Methods for electron-phonon systems\label{subsec:phonons}}

A significant limitation of DMRG and exact diagonalizations
is that they require a finite basis for each site.
In electron-phonon lattice models such as the Holstein-Hubbard
model~(\ref{holstein}), the number of
phonons (bosons) is not conserved and the Hilbert space is infinite
for each site representing an oscillator.
Of course, the number of phonons can be artificially constrained
to a finite number $M$ per site, but 
the number $M$ needed for an accurate
treatment may be quite large. 
This often severely constrains the system size or the
regime of coupling which may be studied with 
DMRG~\cite{Samuel} and exact diagonalizations
(sect.~\ref{sec:ed}). 

Here, we describe two methods for treating systems including sites
with a large Hilbert space.
Both methods use the information contained in a density matrix
to reduce the computational effort required for the study of 
such systems.
The first method~\cite{pseudo}, called pseudo-site method,
is just a modification of the usual DMRG technique 
which allows us to deal more efficiently with sites having a large
Hilbert space.
The second method~\cite{optimal,optimal2}, 
called the optimal basis method,
is a procedure for generating a controlled
truncation of the phonon Hilbert space, which allows the use of a
very small optimal basis without significant loss of accuracy.

\subsubsection{Pseudo-site method\label{subsubsec:pseudo}}

The DMRG algorithms presented above can easily be generalized to treat
systems including phonons (bosons).
In a standard implementation of the DMRG method, however,
each boson forms one lattice site 
and thus memory and CPU time requirements increase
as $M^2$ and $M^3$, respectively. 
Therefore, performing calculations for the Holstein model
requires much more computer resources 
than computations for purely electronic systems.

To understand the basis of the pseudo-site
approach~\cite{pseudo},
it is important to note that, in principle, the computer 
resources 
required by DMRG increase linearly with the number of lattice 
sites.
Thus, DMRG is much better able to handle several 
few-state sites rather than one many-state site.
The key idea of the pseudo-site approach is to 
transform
each boson site with $M=2^{P}$ states into $P$ pseudo-sites
with 2 states.
This approach is motivated by a familiar concept:
The representation of a number in binary form.
In this case the number is the boson state index $s$ going from
0 to $P-1$.
Each binary digit $r_j$ is represented by a pseudo-site, which can 
be occupied ($r_j=1$) or empty ($r_j=0$).
One can think of these pseudo-sites as hard-core bosons.
Thus, 
the level (boson state) with index $s=0$ is represented by $P$ empty
pseudo-sites, while the highest level, $s=2^{P}-1$,
is represented by one boson on each of the $P$ pseudo-sites.

\begin{figure}
\begin{center}
\includegraphics[width=4.5cm]{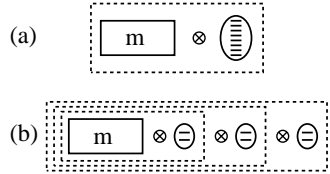}
\caption{
Symbolic representations (a) of the standard DMRG approach 
for $M=8$ 
and (b) of the pseudo-site approach for $P=3$.
}
\label{fig:pseudo}
\end{center}
\end{figure}

Figure~\ref{fig:pseudo} illustrates the differences between
standard and pseudo-site DMRG approaches for $M=8$ ($P=3$).
In the standard approach [fig.~\ref{fig:pseudo}(a)],
a new block (dashed rectangle) is built up
by adding a boson site (oval)
with $M$ states to another block (solid rectangle) 
with $m$ states.
Initially, the Hilbert space of the new block
contains $m M$ states and
is truncated to $m$ states according to the DMRG
method.
In the pseudo-site approach [fig.~\ref{fig:pseudo}(b)],
a new block is made of the previous block with $m$ states and 
one pseudo-site with two states.
The Hilbert space of this new block
contains only $2 m$ states and is also
truncated to $m$ states according to the DMRG
method.
It takes $P$ steps to make the final block 
(largest dashed rectangle)
including the initial block and all pseudo-sites,
which is equivalent to the new block 
in fig.~\ref{fig:pseudo}(a).
However, at each step we have to manipulate
only a fraction $2/M$ of the bosonic Hilbert space.

To implement this pseudo-site method,
we introduce $P$ pseudo-sites $j=1,...,P$
with a two-dimensional Hilbert
space $\{|r_j \rangle, r_j = 0,1 \}$ and the operators
$a^\dag_j, a^{\phantom{\dag}}_j$  such that
$ a^{\phantom{\dag}}_j |1\rangle = |0 \rangle,
\ a^{\phantom{\dag}}_j |0\rangle = 0$
and $a^\dag_j$ is the hermitian conjugate of $a^{\phantom{\dag}}_j$.
These pseudo-site operators have the same properties
as hard-core boson operators: 
$a^{\phantom{\dag}}_j a^\dag_j + a^\dag_j a^{\phantom{\dag}}_j = 1$,
and operators on different pseudo-sites commute.
The one-to-one mapping
between a boson level $|s\rangle$, $s = 0,...,M-1$,
where $b^\dag b |s\rangle = s |s\rangle$,
and the $P$-pseudo-site
state $|r_1, r_2, ..., r_P \rangle$
is given by the relation
$s = \sum_{j=1}^{P}  2^{j-1} r_j $
between an integer number and its binary representation.
The next step is to write all boson operators in terms  of
pseudo-site operators.
It is obvious that the boson number operator is
given by
$N_{\rm b}  = b^\dag b  = \sum_{j=1}^{P}
 2^{j-1} \, a^\dag_j a^{\phantom{\dag}}_j$.               
Other boson operators take a more complicated form.
For instance, to calculate the representation of $b^\dag$ we
first write $b^\dag = B^\dag  \sqrt{N_{\rm b}+1}$,
where $B^\dag |s\rangle = |s+1\rangle$.
The pseudo-site operator representation of the second term is
\begin{equation}
\sqrt{N_{\rm b}+1} =
\sum_{s=0}^{M-1} \sqrt{s+1} \
 A_1(r_1) \, A_2(r_2) ... A_P(r_P)  ,
\end{equation}
where $A_j(1) = a^\dag_j a^{\phantom{\dag}}_j$, 
$A_j(0) = a^{\phantom{\dag}}_j a^\dag_j$ and
the $r_j$ ($j=1,..,P$) are given by the binary digits of $s$.
For $B^\dag$ we find
\begin{equation}
B^\dag   = 
a^\dag_1 + a^\dag_2  a^{\phantom{\dag}}_1 
+ a^\dag_3  a^{\phantom{\dag}}_2  a^{\phantom{\dag}}_1 + ...
+ a^\dag_P  a^{\phantom{\dag}}_{P-1}  
a^{\phantom{\dag}}_{P-2}  ... a^{\phantom{\dag}}_1 .
\end{equation}
Thus one can substitute $P=\log_2(M)$ pseudo-sites for each 
boson site in the lattice and rewrite the system Hamiltonian 
and other operators
in terms of the pseudo-site operators.
Then the finite system DMRG algorithm 
can be used to
calculate the properties of this system of interacting
electrons and hard-core bosons.

The pseudo-site approach outperforms the standard approach
when computations become challenging.
For $M = 32$, the pseudo-site approach is already faster 
than the standard approach by two orders of magnitude.
With the pseudo-site method it is possible
to carry out calculations on lattices large enough 
to eliminate finite size effects
while keeping enough states per phonon mode 
to render the phonon Hilbert space truncation errors negligible. 
For instance, this technique provides some of the most accurate 
results~\cite{pseudo,Romero} 
for the polaron problem in the Holstein model 
(see also the paper by Fehske, Alvermann, Hohenadler and Wellein).
It has also been successfully used to study quantum phase transitions
in the 1D half-filled Holstein-Hubbard model
(see ref.~\cite{peierls} and
the separate paper by Fehske and Jeckelmann).

\subsubsection{Optimal phonon basis\label{subsubsec:optimal}}

The number of phonon levels $M$ needed for an accurate treatment
of a phonon mode can be strongly reduced by 
choosing a basis which minimizes the error due to the truncation
of the phonon Hilbert space instead of using
the bare phonon basis made of the lowest eigenstates of the operators 
$b^\dag_i b^{\phantom{\dag}}_i$.
As with DMRG,
in order to eliminate phonon states without loss
of accuracy, one should transform to the basis of 
eigenvectors of the reduced density matrix 
and discard states with low probability.
The key difference is that here the subsystem is a single site.
To be specific, consider the translationally invariant
Holstein-Hubbard model~(\ref{holstein}).
A site includes both the phonon levels and the 
electron degrees of freedom.
Let $\alpha$ label the four possible electronic states 
of a particular
site and let $s$ label the phonon levels of this site.
Let $j$ label the combined states of all of the rest of the
sites. 
Then a wave function of the system can be written as
\begin{equation}
|\psi\rangle = \sum_{\alpha,s,j} \psi_{\alpha s,j}
| \alpha,s\rangle | j \rangle .
\label{eq:wfn}
\end{equation}
The density matrix for this site for a given 
electronic state
$\alpha$ of the site is 
\begin{equation}
\rho^\alpha_{s,r}
    = \sum_{j} \psi_{\alpha s,j} \psi_{\alpha r,j}^{*}  ,
\label{eq:rho}
\end{equation}
where $r$ also labels the phonon levels of this site.
Let $w_{\alpha k}$ be the eigenvalues  and $\phi_{\alpha k}(n)$
the eigenvectors of $\rho$, 
where $k$ labels the different eigenstates for a given
electronic state of the site. 
The $w_{\alpha k}$ are the probabilities of 
the states $\phi_{\alpha k}$
if the system is in the state~(\ref{eq:wfn}). 
If $w_{\alpha k}$ is negligible, the
corresponding eigenvector can be discarded from the
basis for the site without affecting the state~(\ref{eq:wfn}). 
If one wishes to keep a limited number of states $m$ for a site, 
the best states to keep are the eigenstates of the
density matrices~(\ref{eq:rho}) with the largest eigenvalues. 
In EP  
systems, these $m$ eigenstates form an optimal 
phonon basis\index{optimal phonon basis}.
We note that we obtain different optimal phonon states 
for each of 
the four electron states of the site. 

\begin{figure}[t]
\begin{center}
\includegraphics[width=3.2cm]{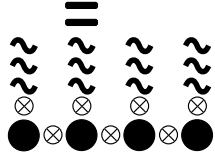}
\caption{Schematic representation of  the big site algorithm.
Each site (circles) has the electronic degrees of freedom
and three optimal states (wiggly bars).
The big site (second site from the left) has the optimal states
plus two bare levels (straight bars).
\label{fig:optimal}
}
\end{center}
\end{figure}

Unfortunately, in order to obtain the optimal phonon states, 
we need
the target state~(\ref{eq:wfn}), which we do not know 
--usually we want the optimal states to help get this state.
This problem can be circumvented in several ways~\cite{optimal}.
Here, we describe one algorithm in conjunction
with an exact diagonalization approach (sect.~\ref{sec:ed})
but it can also be incorporated into a standard DMRG algorithm. 
One site of the system (called the big site) has both optimal
states and a few extra phonon levels. 
(These $n$ extra levels are taken from a set of $M \gg m$ bare 
levels but are explicitly orthogonalized to the current optimal states.)
To be able to perform an exact diagonalization of the system,
each site of the lattice is allowed to have only a small number of 
optimal phonon levels, $m \sim 3 - 4$. 
This approach is illustrated in fig.~\ref{fig:optimal}. 
The ground state of the Hamiltonian is calculated
in this reduced Hilbert space using an exact diagonalization technique.
Then the density matrix (\ref{eq:rho}) of the big site 
is diagonalized. 
The most probable $m$ eigenstates are new optimal phonon states,
which  are used on all other sites for the next diagonalization. 
Diagonalizations must be repeated until the optimal states have
converged. 
Each time different extra phonon levels are used for 
the big site.
They allow improvements of the optimal states by mixing 
in the $M$ bare states little by little.

The improvement coming from using optimal phonon states
instead of bare phonon levels is remarkable.
For the Holstein-Hubbard model\index{Holstein model}
the ground state energy converges very rapidly as a function
of the number of optimal phonon levels~\cite{optimal,optimal2}.
Two or three optimal phonon states per site 
can give results as accurate as with a hundred or more bare 
phonon states per site. 
For intermediate coupling ($\omega_{0}=t$, $g=1.5$, and $U=0$) in the
half-filled band case, the energy is accurate to 
less than 0.1\% using only two optimal levels,
whereas keeping eleven bare levels the error is still greater 
than 5\%.

Combined with ED techniques the above algorithm for generating optimal
phonon states can be significantly improved~\cite{optimal3,Weisse}.
First, the use of a different phonon basis for the big site
artificially breaks the system symmetries.  
This can be solved by including all those states into the
phonon basis that can be created by symmetry operations 
and by summing the density matrices generated with respect to every site.
Second, the effective phonon Hilbert space is unnecessarily large
if one uses the configuration shown in fig.~\ref{fig:optimal}.
As eigenvalues $w_{\alpha k}$ of the density matrix decrease very
rapidly we can introduce a cut-off for the lattice phonon states,
which is reminiscent of the energy or phonon number cut-off~(\ref{eq:cutoff})
discussed in subsubsect.~\ref{subsubsec:hst}, to further reduce the phonon Hilbert space
dimension without loss of accuracy.

In the Holstein-Hubbard model it is possible to transfer optimal 
phonon states from small systems to larger ones because of 
the localized nature of the phonon modes.
Therefore, one can first calculate a large optimal phonon basis
in a two-site system and then use it instead of the bare 
phonon basis to start calculations on larger lattices.
This is the simplest approach for combining the optimal
phonon basis method with standard DMRG techniques for large
lattices such as the infinite- and finite-system 
algorithms~\cite{Friedman}.

The features of the optimal phonon states can sometimes be understood
qualitatively~\cite{optimal,optimal2}.
In the weak-coupling regime optimal states are simply
eigenstates of an oscillator with an equilibrium position
$\langle q \rangle \approx 2g$ as predicted by 
a mean-field approximation.
In the strong-coupling regime ($g^2 \omega_0 \gg U,t$)
the most important optimal phonon states
for $n_i=0$ or 2 electrons on the site can be obtained
by a unitary Lang-Firsov transformation of the bare phonon ground state
$S(g) = e^{-g \sum_i (b_i^\dag - b^{\phantom{\dag}}_i) n_i}$,
in agreement with the strong-coupling theory.
The optimal phonon state for a singly occupied site 
is not given by the Lang-Firsov transformation but 
is approximately the superposition of the optimal phonon states
for empty or doubly occupied sites due to retardation 
effects~\cite{optimal}.
In principle, the understanding gained from an analysis
of the optimal phonon states calculated numerically
with our method could be used
to improve the optimized basis used in variational approaches
(see the related paper by Cataudella).

An interesting feature of the optimal basis approach
is that it provides a natural way to dress electrons 
locally with phonons~\cite{optimal2}.
This allows us to define creation and annihilation operators 
for composite electron-phonon objects like small polarons and bipolarons
and thus to calculate their spectral functions.
However, the dressing by phonons at a finite distance 
from the electrons is completely neglected with this method.

\section{Dynamical DMRG \label{sec:dynamical}}

Calculating the dynamics of quantum many-body systems
such as solids has been a long-standing problem of theoretical physics
because many experimental techniques probe the low-energy, 
low-temperature dynamical properties of these systems.  
For instance, spectroscopy experiments, such as
optical absorption, photoemission, or nuclear magnetic resonance,
can measure dynamical correlations between an external 
perturbation and the response of electrons and phonons
in solids~\cite{Kuzmany}.

The DMRG method has proved to be extremely accurate for 
calculating the properties of very large
low-dimensional correlated systems and even  
allows us to investigate static properties in the thermodynamic limit
The calculation of high-energy excitations and dynamical
spectra for large systems has proved to be more difficult 
and has become possible only recently with the 
development of the dynamical DMRG (DDMRG) approach~\cite{ddmrg0,ddmrg}. 
Here we first discuss the difficulty in calculating excited
states and dynamical properties within the DMRG approach and 
several techniques which have been developed for this
purpose. Then we present the DDMRG method and a
finite-size scaling analysis for dynamical spectra.
In the final section,
we discuss the application of DDMRG  
to electron-phonon systems.  

\subsection{Calculation of excited states and dynamical properties\label{subsec:dynamics}}

\begin{figure}[tb]
\begin{center}
\includegraphics[width=6.0cm]{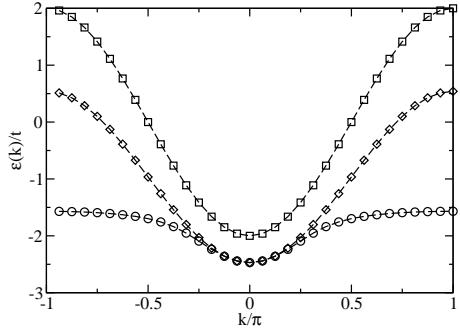}
\caption{Dispersion of the lowest 32 eigenenergies in the 
one-dimensional Holstein model with one electron on
a 32-site lattice for
$\omega_{0}=t$ and $g=1$ calculated with
DMRG (circles).
The diamonds show a hypothetical polaron dispersion
$\epsilon(k)=E_{\rm p}-2t^*[\cos(k)-1]$, where 
$E_{\rm p}=-2.471t$ and the effective hopping 
$t^*=0.753t$ have been fitted to
the DMRG dispersion around $k=0$.
The squares show the free-electron dispersion
$\epsilon(k)=-2t\cos(k)$.
}
\label{fig:polaron}
\end{center}
\end{figure}

The simplest method for computing excited states within DMRG is the
inclusion of the lowest $R$ eigenstates as target 
instead
of the sole ground state, so that the RG transformation
produces effective Hamiltonians describing these states 
accurately.
As an example, fig.~\ref{fig:polaron} shows the dispersion
of the lowest 32 eigenenergies as a function of the momentum
$\hbar k$
in the one-dimensional Holstein model on a 32-site ring
with one electron (the so-called polaron problem,
see the paper by Fehske, Alvermann, Hohenadler, and Wellein
for a discussion of the polaron physics in that model). 
These energies have been calculated
with the pseudo-site DMRG method 
using the corresponding 32 eigenstates as targets. 
Unfortunately, this approach is limited to small number $R$ of targets
(of the order of a few tens), which is not sufficient
for calculating a complete excitation spectrum.  

\subsubsection{Dynamical correlation functions}

The (zero-temperature) 
linear response of a quantum system to a time-dependent
perturbation is often given by dynamical correlation functions
(with $\hbar=1$)
\begin{equation}
G_{A}(\omega + i \eta) = - \frac{1}{\pi}
\langle \psi_0| A^{\dag} \frac{1}{E_0+\omega + i \eta - H} A
|\psi_0\rangle ,
\label{dynamCF}
\end{equation}
where $H$ is the time-independent Hamiltonian of the
system, $E_0$ and $|\psi_0 \rangle$ are its ground-state energy and
wavefunction, $A$ is the quantum operator corresponding to the physical
quantity which is analyzed, and $A^{\dag}$ is the Hermitian conjugate
of $A$. 
A small real number $\eta$ is used in the calculation to
shift the poles of the correlation function into the complex plane.
As a first example, the real part $\sigma_1(\omega > 0)$ 
of the optical conductivity 
is related to the imaginary part of the dynamical
current-current correlation function, which corresponds
to the operator
$                  
A =  {it} \sum_{j\sigma}
(c^\dag_{j+1\sigma} c^{\phantom{\dag}}_{j\sigma} - 
c^\dag_{j\sigma} c^{\phantom{\dag}}_{j+1\sigma})
$                        
in the above equation for the 1D Hubbard and Holstein-Hubbard
models.
As another example, on a one-dimensional lattice
the angle-resolved photoemission spectrum is related 
to the spectral function $A(k,\omega)$ given by the imaginary
part of (\ref{dynamCF}) with the operator
$A =  \frac{1}{\sqrt{N}} \sum_{j} e^{ikj} c_{j\sigma}$. 

In general, we are interested in the imaginary part
of the correlation function
\begin{equation}
\label{spectral}
I_{A}(\omega + i\eta) = {\rm Im} \ G_{A}(\omega + i \eta) 
 =  \frac{1}{\pi} \langle \psi_0 | A^{\dag}
\frac{\eta}{(E_0+\omega -H)^2 + \eta^2} A |\psi_0 \rangle 
\end{equation}
in the $\eta \rightarrow 0$ limit,
$                 
I_{A}(\omega) = \lim_{\eta \rightarrow 0} I_{A}(\omega + i \eta)
 \geq 0 .
$                 
It should be noted that the spectrum $I_{A}(\omega + i \eta)$ for any
finite $\eta > 0$ is equal to the convolution
of the spectral function  $I_{A}(\omega)$
with a Lorentzian distribution of width $\eta$
\begin{equation}
I_{A}(\omega + i \eta) =   
\int_{-\infty}^{+\infty} d\omega' I_{A}(\omega')
\frac{1}{\pi}\frac{\eta}{(\omega-\omega')^2+\eta^2} 
>  0  .
\label{convolution}
\end{equation}

Let $|n\rangle, n=0,1,2, \dots$ be the complete set of eigenstates
of $H$ with eigenenergies $E_n$  ($|n=0\rangle$ corresponds to
the ground state $|\psi_0\rangle$).
The spectral function~(\ref{spectral}) can be written in the
so-called Lehmann spectral representation 
\begin{equation}
\label{lehmann}
I_{A}(\omega + i \eta) =  
\frac{1}{\pi} \sum_{n} |\langle n|A|0\rangle|^2 
\frac{\eta}{(E_n -E_0-\omega)^2 + \eta^2}  ,
\end{equation}
where
$E_n -E_0$ is the excitation energy and $|\langle n|A|0\rangle|^2$ 
the spectral weight of the $n$-th excited state.
Obviously, only states with a finite spectral weight contribute
to the dynamical correlation function~(\ref{dynamCF})
and play a role in the dynamics of the physical quantity
corresponding to the operator $A$.
In the one-dimensional half-filled Hubbard model
the Hilbert space dimension increases exponentially with the
number of sites $N$ but the number of eigenstates with non-zero
matrix elements $\langle n|A|0\rangle$ increases only as a power-law
of $N$ for the
optical conductivity or the one-electron density of states.

\subsubsection{Symmetries \label{subsubsec:symmetry}}

The simplest method for calculating specific excited states 
uses symmetries of the system.
If symmetry operators are well defined
in every subsystem, DMRG calculations can be carried out 
to obtain the $R$ lowest eigenstates in a specific
symmetry subspace. 
It is also possible to target simultaneously 
the lowest eigenstates in two different subspaces
and to calculate matrix elements $\langle n| A|0 \rangle$ 
between them. 
This allows one to reconstruct the dynamical correlation function 
at low energy using eq.~(\ref{lehmann}).

There are several approaches for using 
symmetries in DMRG calculations. 
From a computational point of view, the best approach is an
explicit implementation of the corresponding conserved quantum numbers 
in the program
This is easily done for so-called additive quantum numbers
such as the particle number or the projection of the total spin
onto an axis. 
For instance, if the  $z$-projection of the 
total spin is conserved 
(i.e., the spin operator $S_z$ commutes with the 
Hamilton operator $H$), one can 
calculate the lowest eigenstates for various quantum numbers $S_z$
to investigate spin excitations~\cite{SteveHuse}.
As another example, if we study a system with $N_{\rm e}$ electrons, we
can compute the ground-state energy $E_0(N')$ for 
different number $N'$ of electrons around $N_{\rm e}$
and thus obtain the (charge) gap
$E_{\rm g1} = E_0(N_{\rm e}+1) + E_0(N_{\rm e}-1) - 2 E_0(N_{\rm e})$   
in the spectrum of free electronic charge excitations
(see the separate paper by Fehske and Jeckelmann
for an application of this approach).  
The extension of DMRG to non-abelian symmetries
such as the $SU(2)$ spin symmetry of the Hubbard model
is presented in ref.~\cite{Ian}.

A second approach for using symmetries 
is the construction of projection matrices 
onto invariant subspaces of the Hamiltonian.  
They can be used to project the matrix representation
of the superblock Hamiltonian~\cite{Ramasesha}
or the initial wave function of the iterative 
algorithm used to diagonalize the superblock Hamiltonian~\cite{Boman}
(so that it converges to eigenstates in the chosen subspace).
This approach has successfully been used
to study optical excitations in one-dimensional
models of conjugated polymers~\cite{Boman,Ramasesha_review}.

A third approach consists in adding an interaction term to the system 
Hamiltonian $H$
to shift the eigenstates with the chosen symmetry 
to lower energies. For instance, if the total spin $\mathbf{S}^2$ 
commutes
with the Hamilton operator $H$, one applies the DMRG method
to the Hamiltonian $H^{\prime} = H + \lambda \mathbf{S}^2$ with
$\lambda > 0$ to obtain the lowest singlet eigenstates
without interference from the $\mathbf{S}^2 > 0$ 
eigenstates~\cite{Stephane}.

Using symmetries is the most efficient and accurate approach  
for calculating specific low-lying excited states with DMRG.
However, its application is obviously restricted to those problems
which have relevant symmetries and it
provides only the lowest $R$ eigenstates
for given symmetries, where $R$
is at most a few tens for realistic applications. Thus this approach
cannot describe high-energy excitations nor any complex or
continuous dynamical spectrum.

\subsubsection{Lanczos-DMRG}

The Lanczos-DMRG approach was introduced by Karen Hallberg in 1995
as a method for studying dynamical properties of lattice quantum
many-body systems~\cite{Hallberg}.
It combines DMRG with the continuous fraction
expansion technique, also called Lanczos algorithm 
(see subsubsect.~\ref{subsubsec:sdm}),  
to compute the dynamical correlation function~(\ref{dynamCF}).
Firstly, the Lanczos algorithm is used to calculate the complete 
dynamical spectrum of a superblock Hamiltonian.
Secondly, some Lanczos vectors are used as DMRG target 
in an attempt at constructing an effective Hamiltonian which 
describes excited states contributing to the correlation function accurately.
Theoretically, one can systematically improve the accuracy using an
increasingly large enough 
number $L$ of Lanczos vectors as targets.
Unfortunately, the method becomes numerically instable for large
$L$ as soon as the DMRG truncation error is finite.
Therefore, Lanczos-DMRG is not an exact numerical method
except for a few special cases. 
The cause of the numerical instability is the tendency of the Lanczos
algorithm to blow up the DMRG truncation errors.  
In practice, only the first few Lanczos vectors
are included as target.
The accuracy of this type of calculation is unknown
and can be very poor.

From a physical point of view, the failure of
the Lanczos-DMRG approach for complex dynamical spectra can be 
understood.  
With this method one attempts to construct a single effective
representation of the Hamiltonian $H$ which describes the relevant
excited states for all excitation energies.
This contradicts the essence of a RG calculation, which is
the construction of an effective representation of a system
at a specific  energy scale by integrating out the other
energy scales.

Nevertheless, Lanczos-DMRG is a relatively
simple and quick method for calculating dynamical properties
within a DMRG approach and it has already been used in
several works~\cite{Uli_review}.
It gives accurate results for systems slightly larger
than those which can be investigated with exact diagonalization
techniques. 
It is also reliable for larger systems with simple discrete spectra 
made of a few peaks.
Moreover, Lanczos-DMRG gives accurate results for the first
few moments of a spectral function and
thus provides us with a simple independent check of the spectral
functions calculated with other methods.

In the context of EP systems
the Lanczos algorithm has been successfully combined with
the optimal phonon basis DMRG method
(see subsubsect.~\ref{subsubsec:optimal}). 
The optical conductivity, single-electron spectral functions
and electron-pair spectral functions have been 
calculated for the 1D Holstein-Hubbard model 
at various band fillings~\cite{optimal2,Chunli}.
Results for these dynamical quantities
agree qualitatively with results obtained by conventional
exact diagonalizations using powerful parallel 
computers (sect.~\ref{sec:ed}).

\subsubsection{Correction vector DMRG}

Using correction vectors to calculate dynamical correlation
functions with DMRG was first proposed by 
Ramasesha {\it et al.}~\cite{Pati}.
The correction vector associated with $G_A(\omega + i \eta)$ is defined
by
\begin{equation}
\label{CV}
|\psi_A(\omega + i \eta) \rangle = \frac{1}{E_0+\omega + i \eta - H}
| A \rangle ,
\end{equation}
where $| A \rangle = A | \psi_0 \rangle$ is the first
Lanczos vector.
If the correction vector is known, the dynamical correlation
function can be calculated directly
\begin{equation}
G_A(\omega + i \eta) =
-\frac{1}{\pi} \langle A|\psi_A(\omega + i \eta) \rangle .
\label{dynamCF2}
\end{equation}
To calculate a correction vector, one first solves
an inhomogeneous linear equation
\begin{equation}
\left [ (E_0+\omega-H)^2+\eta^2 \right ] | \psi \rangle
= -\eta | A \rangle ,
\label{CVequation1}
\end{equation}
which always has a unique solution
$| \psi \rangle = | Y_A(\omega + i \eta) \rangle$ for $\eta \neq 0$.
The correction vector is then given by
$
|\psi_A(\omega + i \eta) \rangle = | X_A(\omega + i \eta) \rangle
+ i | Y_A(\omega + i \eta) \rangle
$, 
with
\begin{equation}
| X_A(\omega + i \eta) \rangle =
\frac{H-E_0-\omega}{\eta} | Y_A(\omega + i \eta) \rangle  .
\label{CVequation2}
\end{equation}
One should note that the states $| X_A(\omega + i \eta) \rangle$
and $| Y_A(\omega + i \eta) \rangle$ are complex if the state
$|A\rangle$ is not real, but they  always determine the real part and
imaginary part of the dynamical correlation function
$G_A(\omega + i \eta)$, respectively.
The approach can be extended to higher-order dynamic response 
functions such as third-order optical polarizabilities~\cite{Pati} 
and to derivatives of dynamical correlation functions~\cite{ddmrg}.

The distinct characteristic of a correction vector approach 
is that a specific
quantum state~(\ref{CV}) is constructed to compute 
the dynamical correlation function~(\ref{dynamCF}) at each frequency
$\omega$.  
To obtain a complete dynamical spectrum, the procedure has to be
repeated for many different frequencies.
Therefore,
this approach is generally less efficient than
the iterative methods presented in sect.~\ref{sec:sp}
in the context of exact diagonalizations.
For DMRG calculations, however, this is a highly favorable
characteristic.  
The dynamical correlation function can be determined for
each frequency $\omega$ separately using effective representations
of the system Hamiltonian $H$ and operator $A$ which have to
describe a single energy scale accurately.

K\"{u}hner and White~\cite{Till}
have devised a correction vector DMRG method
which uses this characteristic to 
perform accurate calculations 
of spectral functions for all frequencies
in large lattice quantum many-body systems.
In their method, two correction vectors with close frequencies 
$\omega_1$ and $\omega_2$ and finite broadening 
$\eta \sim \omega_2 -\omega_1 >0$
are included as target. 
The spectrum is then calculated in the frequency interval 
$\omega_1 \alt \omega \alt \omega_2$
using eq.~(\ref{dynamCF2}) or the continuous fraction expansion.
The calculation is repeated for several (overlapping)
intervals to determine the spectral function over a large 
frequency range.
This procedure
makes the accurate computation of complex or continuous spectra possible.
Nevertheless, there have been relatively few applications of
the correction vector DMRG method~\cite{Uli_review} because it
requires substantial computational resources and is difficult
to use efficiently.

\subsection{Dynamical DMRG method\label{subsec:ddmrg}}

The capability of the correction vector DMRG method to calculate
continuous spectra shows that using specific target states
for each frequency is the right approach.   
Nevertheless, it is highly desirable to simplify  
this approach and to improve its performance. 
A closer analysis shows that the complete problem of 
calculating dynamical properties can be 
formulated as a minimization problem. 
This leads to the definition of a more efficient and simpler 
method, the dynamical DMRG (DDMRG)~\cite{ddmrg}. 

DDMRG enables accurate calculations
of dynamical properties for all frequencies in large
systems with up to
a few hundred particles using a workstation.
Combined with a proper finite-size-scaling analysis 
(subsect.~\ref{subsec:scaling})
it  also enables the investigation of spectral functions
in the thermodynamic limit.
Therefore, DDMRG provides a powerful new approach for investigating
the dynamical properties
of quantum many-body systems.

\subsubsection{Variational principle}

In the correction vector DMRG method the most time-consuming task is 
the calculation of correction vectors in the superblock
from eq.~(\ref{CVequation1}).
A well-established approach for solving an inhomogeneous linear
equation~(\ref{CVequation1}) is to formulate it as a minimization
problem.
Consider the equation  $M \mathbf{x} = \mathbf{a}$, where
$M$ is a positive definite symmetric matrix with a non-degenerate
lowest eigenvalue, $\mathbf{a}$ is a known vector, 
and $\mathbf{x}$ is the unknown vector to be calculated. 
One can define the function
$W(\mathbf{x})= \mathbf{x} \cdot M \mathbf{x} 
- \mathbf{x} \cdot \mathbf{a}
- \mathbf{a} \cdot \mathbf{x}$,
which has a non-degenerate minimum for the vector 
$\mathbf{x}_{\rm min} = M^{-1} \mathbf{a}$              
which is solution of the inhomogeneous linear equation. 
(K\"{u}hner and White~\cite{Till} used a conjugate gradient 
method to solve this minimization problem.)

Generalizing this idea one can formulate a variational 
principle for dynamical correlation functions.
One considers the functional
\begin{equation}
W_{A,\eta}(\omega, \psi)  = 
\langle \psi | (E_0+\omega-H)^2+\eta^2  | \psi \rangle
+ \eta \langle A | \psi \rangle + \eta \langle \psi | A \rangle .
\label{functional}
\end{equation}
For any $\eta \neq 0$ and a fixed frequency $\omega$ this functional
has a well-defined and non-degenerate minimum
for the quantum state which is solution of eq.~(\ref{CVequation1}), i.e.
$
| \psi_{{\rm min}} \rangle = | Y_A(\omega + i \eta) \rangle  .
$                
It is easy to show that the value of the minimum is related
to the imaginary part of the dynamical correlation function
\begin{equation}
W_{A,\eta}(\omega, \psi_{{\rm min}}) =
-\pi\eta I_A(\omega + i \eta).
\label{minimum}
\end{equation}
Therefore, the calculation of spectral functions can be formulated
as a minimization problem.
To determine $I_A(\omega + i \eta)$ at any frequency $\omega$ and for
any $\eta > 0$, one minimizes the corresponding functional
$W_{A,\eta}(\omega, \psi)$.
Once this minimization has been carried out, the real part of
the correlation function $G_A(\omega + i \eta)$ can be calculated using
eqs.~(\ref{dynamCF2}) and~(\ref{CVequation2}).
This is the variational principle for dynamical correlation functions.
It is clear that if we can calculate $| Y_A(\omega + i \eta) \rangle$
exactly, this variational formulation is completely equivalent to a 
correction vector approach.
However, if we can only calculate an approximate solution with an error
of the order $\epsilon \ll 1$, $|\psi\rangle = | Y_A(\omega + i \eta)
\rangle + \epsilon |\phi\rangle$ with $\langle \phi|\phi \rangle=1$,
the variational formulation is more accurate.
In the correction vector method the error in the spectrum
$I_A(\omega + i \eta)$ calculated
with eq.~(\ref{dynamCF2}) is also of the order of $\epsilon$.
In the variational approach it is easy to show that the error
in the value of the minimum $W_{A,\eta}(\omega, \psi_{{\rm min}})$,
and thus in $I_A(\omega + i \eta)$, is of the order of $\epsilon^2$.
With both methods the error in the real part of $G_A(\omega + i \eta)$
is of the order of $\epsilon$.

The DMRG procedure used to minimize the energy functional 
$E(\psi)$
(see sect.~\ref{sec:dmrg}) can also be used to minimize the
functional $W_{A,\eta}(\omega,\psi)$ and thus to calculate the
dynamical correlation function $G_A(\omega+i\eta)$.
This approach is called the dynamical DMRG method.
In principle, it is equivalent to the correction vector DMRG
method of K\"{u}hner and White~\cite{Till} 
because the same target states are used to build the
DMRG basis in both methods.  
In practice, however, DDMRG
has the significant advantage over the correction
vector DMRG that errors in $I_A(\omega+i\eta)$
are significantly smaller 
(of the order of $\epsilon^2$ instead of $\epsilon$ as
explained above) as soon as DMRG truncation errors are no
longer negligible.

\subsubsection{DDMRG algorithm}

The minimization of the functional $W_{A,\eta}(\omega,\psi)$ is
easily integrated into the usual DMRG algorithm.
At every step of a DMRG sweep through the system lattice,
a superblock representing the system is built and the following
calculations are performed in the the superblock subspace:
\begin{enumerate}

\item The energy functional $E(\psi)$ is minimized using a standard
iterative algorithm for the eigenvalue problem. This yields the ground
state vector $|\psi_0\rangle$ and its energy $E_0$ in the superblock
subspace.

\item The state $|A \rangle$ is calculated.

\item The functional $W_{A,\eta}(\omega,\psi)$ is minimized using an
iterative minimization algorithm.
This gives the first part of the correction vector
$|Y_{A}(\omega+ i\eta)\rangle$ and the imaginary part
$I_A(\omega + i \eta)$ of the dynamical correlation function
through eq.~(\ref{minimum}).

\item  The second part  $|X_{A}(\omega+ i\eta)\rangle$
of the correction vector is calculated using eq.~(\ref{CVequation2}).

\item
The real part of the dynamical correlation function
can be calculated from eq.~(\ref{dynamCF2}).

\item The four states $|\psi_0\rangle$, $|A \rangle$,
$|Y_{A}(\omega+ i\eta)\rangle$, and $|X_{A}(\omega+ i\eta)\rangle$
are included as target in the density matrix renormalization
to build a new superblock at the next step.

\end{enumerate}
The robust finite-system DMRG algorithm must be used to perform
several sweeps through a lattice of fixed size.
Sweeps are repeated until the procedure has converged  to the
minimum of both functionals $E(\psi)$ and $W_{A,\eta}(\omega,\psi)$.

To obtain the dynamical correlation function
$G_{A}(\omega+ i\eta)$ over a range of frequencies, one has to
repeat this calculation for several frequencies $\omega$.
If the DDMRG calculations are performed independently, the computational
effort is roughly proportional to the number of frequencies.
It is also possible to carry out a DDMRG calculation for
several frequencies simultaneously, including several states
$|X_{A}(\omega+ i\eta)\rangle$ and $|Y_{A}(\omega+ i\eta)\rangle$
with different frequencies $\omega$ as target.
As calculations for different frequencies are essentially
independent, it would be easy and very efficient to use  
parallel computers.


Because of the variational principle one naively expects that the DDMRG
results for $I_A(\omega +i\eta)$ must converge monotonically from below
to the exact result as the number $m$ of density matrix eigenstates is
increased.
In practice, the convergence is less regular because of two
approximations made to calculate the functional
$W_{A,\eta}(\omega,\psi)$. 
First, the ground-state energy $E_0$ and the state $|A\rangle$ used
in the definition~(\ref{functional}) of $W_{A,\eta}(\omega,\psi)$
are not known exactly but calculated with DMRG.
Second, one calculates an effective representation of $H$ only 
and assumes that $(H^2)_{\rm eff} \approx (H_{\rm eff})^2$
to compute $W_{A,\eta}(\omega,\psi)$ in the superblock subspace.
These approximations can cause a violation of the variational bound
$W_{A,\eta}(\omega,\psi) \geq -\pi \eta I_A(\omega +i\eta)$.
In practice, for a sufficiently large number $m$ of
density matrix eigenstates kept, 
the absolute errors in $I_A(\omega +i \eta)$
decrease systematically with increasing $m$.
Errors becomes negligible if enough states are kept to 
make the discarded weight vanish.  
It is also possible to estimate the accuracy of a DDMRG
calculation from the results obtained for different values of $m$.
Therefore, DDMRG is an exact numerical method as defined in
the introduction (sect.~\ref{sec:intro}).

\begin{figure}[t]
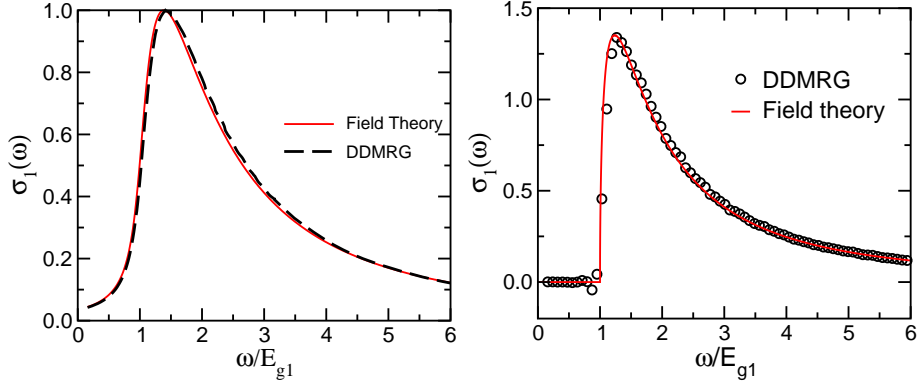

\begin{center}
\includegraphics[width=6.0cm]{jeckel_fehske_fig11a.eps}
\includegraphics[width=6.0cm]{jeckel_fehske_fig11b.eps}
\caption{Optical conductivity of the one-dimensional half-filled
Hubbard model for $U=3t$. Left panel: DDMRG result calculated
on a 128-site lattice using a broadening $\eta=0.1t$ (dashed line) and
field-theoretical prediction for the infinite chain~\cite{ddmrg0}
broadened with a Lorentzian of width $\eta=0.1t$ (solid line).
Right panel: the field-theoretical prediction without broadening
(solid line) and the DDMRG result after deconvolution (circles). 
}
\label{fig:ddmrg}
\end{center}
\end{figure}

The accuracy of the DDMRG approach for 1D
correlated electron systems has been demonstrated by numerous
comparisons with exact analytical 
results~\cite{ddmrg0,ddmrg,Fabian,Eric2,Holger,Holger2,Satoshi}.
As an example, fig.~\ref{fig:ddmrg} shows the optical conductivity
of the 1D half-filled Hubbard model for
$U=3t$. The DDMRG result (calculated using a broadening
$\eta = 0.1t$ on a 128-site lattice) agrees perfectly with
the field-theoretical prediction (also broadened with a 
Lorentzian distribution of width $\eta$)~\cite{ddmrg0,Fabian}.

\subsection{Spectrum in the thermodynamic limit\label{subsec:scaling}}
 
 DDMRG allows us to calculate spectral functions of a large but
 finite system with a broadening
 given by the parameter $\eta > 0$.
 To determine the properties of a dynamical spectrum 
 in the
 thermodynamic limit, one has to analyze the scaling of the corresponding
 spectra $I_{N, \eta}(\omega)$ as a function of the system size $N$
 for vanishing broadening $\eta$
 \begin{equation}
 I(\omega) =  \lim_{\eta \rightarrow 0} \lim_{N \rightarrow \infty}
 I_{N, \eta}(\omega)  .
 \label{inflim}
 \end{equation}
 Computing both limits in this equation from numerical results
 for $I_{N, \eta}(\omega)$ requires a lot of accurate data
 for different values of $\eta$ and $N$
 and can be the source of large extrapolation errors.
 A much better approach is to use a broadening $\eta(N) >0$
 which decreases with increasing $N$ and vanishes in the
 thermodynamic limit~\cite{ddmrg}.
 The dynamical spectrum is then given by
 $                  
 I(\omega)  =  \lim_{N \rightarrow \infty} I_{N, \eta(N)}(\omega) .
 $                
 From the existence of both limits in eq.~(\ref{inflim})
 it can be demonstrated that there exists a minimal broadening
 $\eta_0(N) \geq 0$,
 which converges to zero for $N \rightarrow \infty$, 
 such that this procedure is exact for all
 functions $\eta(N)$ with $\eta(N) > \eta_0(N)$ and
 $\lim_{N \rightarrow \infty} \eta(N) = 0$.

 The function $\eta_0(N)$ depends naturally on the specific problem
 studied  and 
 can also vary for each frequency $\omega$ considered.
 For one-dimensional
 correlated electron systems such as the Hubbard model~(\ref{hubbard}),
one finds empirically that a sufficient condition is
 \begin{equation}
 \eta(N)  = \frac{c}{N} ,
 \label{etacondition}
 \end{equation}
 where the constant $c$ is comparable to the effective
 width of the dynamical spectrum $I(\omega)$, 
 which is finite in such lattice models.
 This condition has a very simple physical interpretation.
 The spectral function $I_{N, \eta}(\omega)$ represents the dynamical
 response of the system over a time period $\sim 1/\eta$ after
 one has started to apply an external force.
 Typically, in a lattice model the spectral width is proportional to the
 velocity of the excitations involved in the system response.
 Thus the condition~(\ref{etacondition}) means that excitations are too
 slow to travel the full length $\sim N$ of the system in the time
 interval $\sim 1/\eta$ and do not ``sense'' that the system is finite.

An additional benefit of a broadening satisfying the
condition~(\ref{etacondition})
is that the finite-system spectrum $I_{N,\eta}(\omega)$ becomes
indistinguishable from the infinite-system spectrum with the same
broadening $\eta$ for relatively small $N$.
Therefore, if one knows a spectral function
$I(\omega)$ for an infinite system, its convolution with a
Lorentzian of width $\eta$ can be compared directly with the numerical
results for the finite-system spectrum $I_{N,\eta}(\omega)$.
This is the approach used in fig.~\ref{fig:ddmrg} (left panel) to compare
DDMRG results for a finite lattice with the field-theoretical
prediction for an infinite chain.

Finally, an approximation for an infinite-system (continuous) spectrum 
can be obtained by solving 
the convolution equation~({\ref{convolution}) numerically 
for $I_A(\omega')$
using the numerical DDMRG data for a finite system on
the left-hand side of this equation~\cite{Satoshi}.
Performing such a deconvolution is a ill-conditioned inverse problem,
which can only be solved approximately using some assumptions
on the spectrum properties like its smoothness.  
Therefore, the accuracy of deconvolved DDMRG spectra is unknown.
In practice, however, one obtains often accurate results
as shown in fig.~\ref{fig:ddmrg} (right panel), 
where a deconvolved DDMRG spectrum
is compared to an exact field-theoretical result.

In summary, the dynamical spectrum of an infinite system can be
determined accurately and efficiently from numerical (DDMRG) data for
finite-system spectra using a finite-size scaling analysis with
a size-dependent broadening $\eta(N)$.

\subsection{Application to electron-phonon problems}

The DDMRG algorithm described in the previous section can be applied
to EP systems such as the Holstein-Hubbard model
without modification~\cite{Georg}.
It can naturally be combined with the special DMRG
techniques for systems with bosonic degrees of freedom which
are described in subsect.~\ref{subsec:phonons}. 
As for ground-state simulations,
these special techniques substantially reduce the computational
cost for calculating dynamical properties
of EP systems.
Nevertheless, applications of DDMRG to the Holstein-Hubbard model 
are significantly more costly than those for comparable purely 
electronic systems such as the Hubbard model.

As an illustration, we compare DDMRG and ED-KPM 
(subsubsect.~\ref{subsubsec:kpm}) results for the spectral
functions of an eight-site spinless Holstein model in 
fig.~\ref{fig:ep}. There is an excellent overall
agreement between both methods. The observable
differences are due to the larger broadening $\eta=0.1t$ used
in the DDMRG simulation. This hides the sharpest details
of the spectrum like the oscillations
which are visible in the KPM spectrum.

\begin{figure}[t]
\begin{center}
\includegraphics[width=8.0cm]{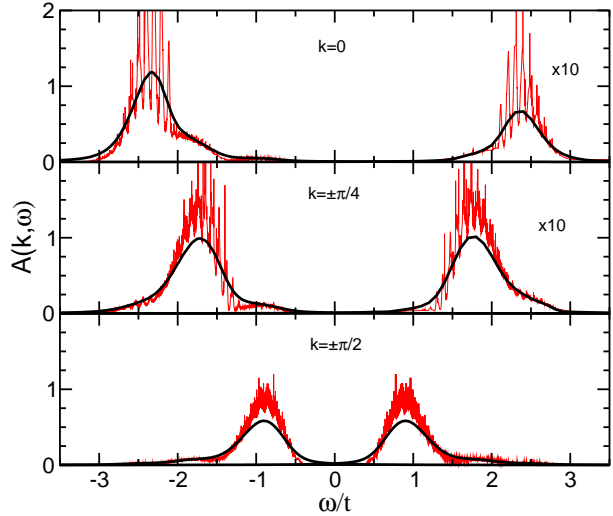}
\caption{
Spectral functions $A(k,\omega)$
(for electron removal, $\omega < 0$, and
electron injection, $\omega > 0$) 
of the spinless Holstein model
at half filling on a 8-site lattice with periodic
boundary conditions.
The system is in the CDW/Peierls insulating phase
($\omega = 0.1 t$ and $g=4$).
The rapidly oscillating thin lines are the KPM results 
while the smooth thick line are the DDMRG results
with the pseudo-site method.
Note that only $|k| \leq \pi/2$ is shown because
$A(k\pm\pi,\omega) = A(k,-\omega)$.
}
\label{fig:ep}
\end{center}
\end{figure}

A significant difference between both methods is the computational
cost. 
These DDMRG calculations took only 150 CPU hours on an Opteron 244
processor (1.8 GHz) and required less than 300 MBytes of memory.
It is thus possible to simulate significantly larger EP
systems than this eight-site lattice with DDMRG while this system
size is the largest one which can be simulated with exact 
diagonalization techniques.

Finally, we note that the broadening $\eta=0.1t$ used for
the DDMRG results in fig.~\ref{fig:ep} is one order of magnitude 
smaller
than the value used for a purely electronic systems of
comparable size (see fig.~\ref{fig:ddmrg}).
It seems that the scaling~(\ref{etacondition}) is not applicable
to EP systems. 
Analytical and numerical investigations will be necessary
to determine the function $\eta_0(N)$ for EP systems
before one can investigate their dynamical properties in the
thermodynamic limit using the techniques described
in subsect.~\ref{subsec:scaling}.

\section{Conclusion}

Exact diagonalization and density matrix renormalization group
techniques are powerful and versatile exact numerical approaches
for investigating the properties of electron-phonon 
lattice models for strongly correlated (low-dimensional)
materials. 
Thanks to recent developments 
we can now calculate dynamical quantities
which are directly related to experimental techniques
used in solid-state spectroscopy and 
often determine the properties in the
thermodynamic limit
using a finite-size scaling analysis.

\appendix

\acknowledgments

We would like to thank B.~B\"auml, H.~Benthien, F.~E\ss ler, 
F.~Gebhard, G.~Hager, S.~Nishimoto, G.~Wellein, and
A.~Weisse for valuable discussions.


\end{document}